\documentclass[journal]{IEEEtran}

\usepackage{url}
\usepackage{xcolor}

\usepackage{hyperref}

\usepackage{latexsym}
\usepackage{amsmath,amsfonts,amssymb}
\usepackage{graphicx}
\usepackage{setspace}
\usepackage{subfigure}
\usepackage{epstopdf}
\usepackage{multicol}
\usepackage{multirow}
\usepackage{amsmath}
\usepackage{algorithm,algorithmic}
\usepackage{xcolor}
\usepackage{booktabs}
\usepackage{amsmath}
\usepackage{adjustbox}
\usepackage{rotating}
\usepackage{relsize}
\usepackage{longtable}
\begin{document}

\title{BlinQS: Blind Quality Scalable Image Compression Algorithm without using PCRD Optimization}

\author{Naveen Cheggoju, ~\IEEEmembership{Member,~IEEE,} and Vishal R. Satpute, ~\IEEEmembership{Member,~IEEE,}
\thanks{Naveen Cheggoju is with School of Electronics (SENSE), VIT AP University, Amaravathi, Vijayawada, Andhra pradesh, 522237, India email: naveen.c@vitap.ac.in }% <-this % stops a space
\thanks{Vishal Satpute is with the Department
of Electronics and Communication Engineering, Visvesvaraya National Institute of Technology, Nagpur, Maharashtra, 440010 India e-mail: vrsatpute@ece.vnit.ac.in}% <-this % stops a space
\thanks{Manuscript received  ,  ; revised  ,  .}}

\maketitle

\begin{abstract}
Quality Scalability is one of the important features of interactive imaging to obtain better perceptual quality at a specified target bit rate. In JPEG 2000, it is achieved using quality layers obtained by Rate-Distortion (R-D) optimization techniques in Tier-II coding. Some important concerns here are: (i) inefficient conventional Post-Compression Rate-Distortion (PCRD) optimization algorithms, (ii) lack of quality scalability for less or single quality layer string. This paper takes the above mentioned concerns into account and proposes a Blind Quality Scalable (BlinQS) algorithm that provides scalability with the least computational complexity. The novel part of this method is to eliminate the Tier-II coding and add a blind string selection algorithm through a normal distribution for efficient rate control. The results obtained suggest that the proposed method achieves better results than JPEG-2000 at single quality layer and achieves results close to JPEG-2000 without using PCRD optimization algorithms. 

\end{abstract}

%
%
%\IEEEpeerreviewmaketitle

\section{Introduction}

\begin{figure}
 \centering
 \includegraphics[width=0.7\columnwidth, height=5cm]{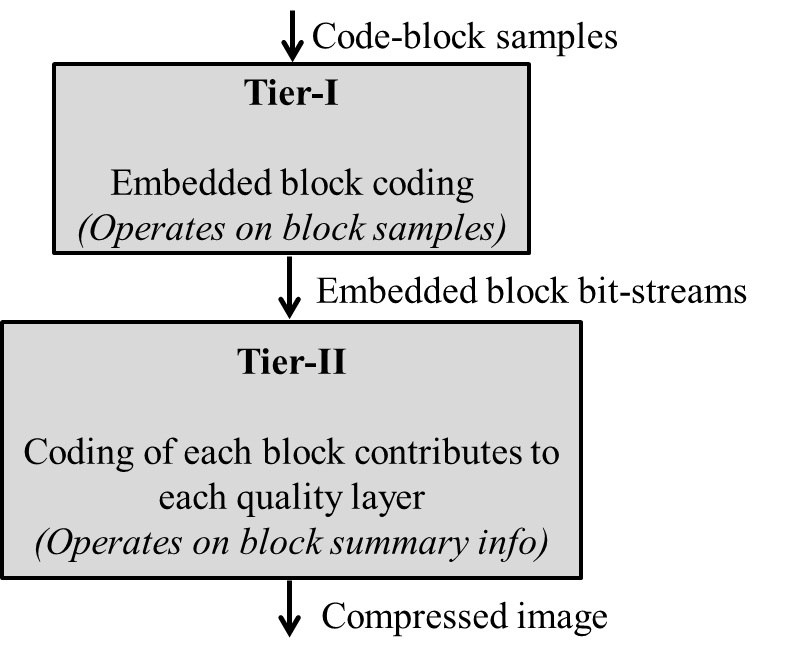}
 \caption{Coding blocks of JPEG-2000}
 \label{fig:jpeg2000_block}
\end{figure}

\IEEEPARstart{S}calability is one of the main features of any interactive device. Scalability may refer to adaptation in size, shape, quality, rate, etc. In the field of image compression, scalability refers to adaptation in resolution, rate, quality and component. Among these, rate scalability and quality scalability need to achieve a good trade-off for maintaining the image quality at a specified or required target rate. This trade-off has been addressed in the new image compression standard JPEG-2000 using the concept of \textit{quality layers} \cite{taubman2000high}, \cite{taubman2002jpeg2000}. These layers are generated in iterative manner using Post-Compression Rate-Distortion (PCRD) optimization algorithms for all the individual code-blocks. To implement  R-D optimization algorithms, JPEG-2000 creates a Tier-II coding system module as shown in Figure \ref{fig:jpeg2000_block} \cite{taubman2000high}. This module takes the block summary of each code-block as the inputs and arranges them in increasing order of  quality for the specified target rates. Some important concerns of this coding are: (i) inefficient conventional  Post-Compression Rate-Distortion (PCRD) optimization algorithms (iterative in nature), (ii) lack of quality scalability for less or single quality layer string. Research communities around the globe have been carrying out research to address these key issues in scalable compression. This has been the prime obstacle of JPEG-2000 widespread adoption in entertainment and broadcast sectors \cite{taubmanhigh}.

For interactive imaging in entertainment and broadcast sectors, scalability refers to two targets: (i) low computational complexity for fast processing, and, (ii) highly independent data rearrangement to achieve optimal quality at any required rate. JPEG-2000 has attempted to achieve this and earned more popularity in the field of interactive imaging, especially in the cloud-based content distribution applications. Having earned all the praises for its flexibility, JPEG-2000 has its own limitations pertaining to the computational complexity of the block coding (PCRD) algorithms. This has been the prime obstacle of JPEG-2000 widespread adoption in entertainment and broadcast sectors \cite{taubmanhigh}. To break through the obstacles and make JPEG-2000 more flexible researchers throughout the world are continuing their research by keeping the following as the prime targets: 
\begin{enumerate}
 \item low computational complexity for fast processing, and,
 \item highly independent data rearrangement to achieve optimal quality at any required rate.
\end{enumerate}

To achieve scalability at low computational complexity, choosing effective optimization algorithm is necessary. This challange has been taken up by \cite{1377377}, \cite{1626299}, \cite{1362522}, \cite{chebil2006pre} by coding the data and obtain the rate-disortion simultaneously . However, these algorithms fail to decrease the computational load on the encoder. To further reduce the computational complexity, wavelet data based and step size based rate-distortion algorithms have been presented in \cite{parisot2002high} and \cite{4099394} respectively. Later, other approaches including Lagrange multiplier have been proposed in \cite{1416230}, \cite{4107067}. In \cite{yeung2005efficient}, authors have proposed three rate control methods (successive bit-plane rate allocation (SBRA), priority scanning rate allocation
(PSRA) and priority scanning with optimal truncation (PSOT)) over PCRD algorithms for reducing computational complexity and memory usage. These rate control methods have provided different trade-off among quality, complexity, memory utilization and coding delay. In \cite{etesaminia2017efficient}, a low complexity R-D optimization method based on a reverse order for resolution levels and coding passes has been proposed. This method has attained a comparable quality performance with the conventional method, maintaining low complexity. These algorithms have proven to be efficient, but, high throughput has not been achieved. In \cite{taubmanhigh}, D. Taubman et. al, have coined the term FBCOT (Fast Block Coding with Optimized Truncation), to widespread the adoption of JPEG-2000 in entertainment and broadcast sectors. This JPEG-2000 compatible algorithm has sole target of increasing the throughput by reducing the computational complexity. This algorithm offers 10X or higher speed compared to the previous algorithms with a slight sacrifice in the coding efficiency. In order to optimize JPEG-2000 for image transmission through wireless networks, Joint Source-Channel Coding (JSCC) has been proposed in \cite{bi2017joint} and congestion control for interactive applications over SDN networks has been proposed in \cite{naman2018responsive}. Further detailed study of rate-distortion optimization for JPEG-2000 is found in \cite{ortega1998rate}, \cite{auli2007model}. These R-D optimization algorithms have been significant in reducing the computational complexity but the problem of scalability for any required rate remains unsolved. 

To solve this problem of scalability for any bit rate, there is a need to develop algorithms which do not rely on the rate-distortion algorithms at the encoder, rather calculate the quality layers at the transcoder without actually having the knowledge of the code-block information. One such method has been proposed in \cite{auli2008jpeg2000}, in which characterisation of code-blocks does not depend on the distortion measures related to the original image. The method proposed in \cite{auli2008jpeg2000}, has been inspired by the algorithms presented in \cite{auli2006efficient}, \cite{auli2007low}, which also speak about achieving the better quality of reconstruction when there is a compressed string with single quality layer or less number of quality layers. As this method is computationally expensive another method called Block-Wise Layer Truncation (BWLT) has been proposed in \cite{auli2010enhanced}. The main insight behind BWLT is to dismantle and reassemble the to-be-fragmented layer by selecting the most relevant codestream segments of codeblocks within that layer. All these methods are targetted to achieve optimum scalability for single layered or less number of quality layered strings. In \cite{auli2009distortion} and \cite{auli2011scanning}, authors focussed on proposing new estimators to approximate the distortion produced by the successive coding of transform coefficients in bitplane image coders. Recently CNN based lossy image compression with multiple bit-rate has been proposed in \cite{cai2020learning}. This paper focusses on learning multiple bit-rates from a single CNN using Tuceker Decomposition Network (TDNet). Even using CNN based approach, the optimum quality is achieved only for the predefined bit rates learnt at the encoder. 

Investigation has been done in modification of bit plane strategies using several theoretical-practical mechanisms conceived from rate-distortion theory. The research work presented in this paper, is mainly focussed on achieving the scalability for even a single-layered string with the minimal complexity. To achieve this, a strong decision making is necessary at the transcoder, to optimally choose the code-blocks and the truncation points. 

Further the paper is organized in the following order: Section-\ref{sec:qualityscaling_JPEG2000} discusses R-D optimizaion and quality scalable algorithm used in JPEG-2000, Section-\ref{sec:proposed_blindquality_method} introduces the proposed method for achieving Blind Quality Scalable Image compression, Section-\ref{sec:results} presents comparative analysis of BlinQS with the JPEG-2000 standard and finally, Section-\ref{sec:conclusion} draws the conclusion of the work followed by the references. 

\section{Quality scaling in JPEG-2000} \label{sec:qualityscaling_JPEG2000}
\begin{figure}[h]
 \centering
 \includegraphics[width=0.9\columnwidth, height=7cm]{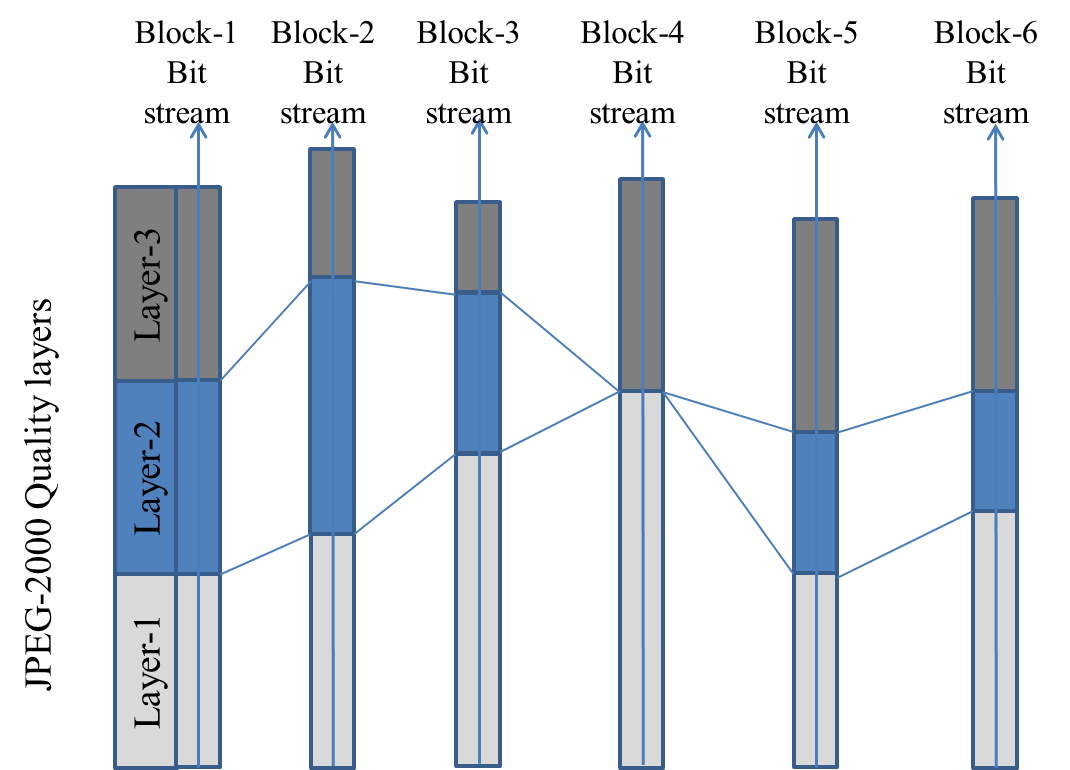}
 \caption{Illustration of Quality Layers in JPEG-2000}
 \label{fig:jpeg2000_qualitylayers}
\end{figure}

\begin{figure}[h]
 \centering
 \includegraphics[width=0.9\columnwidth, height=5cm]{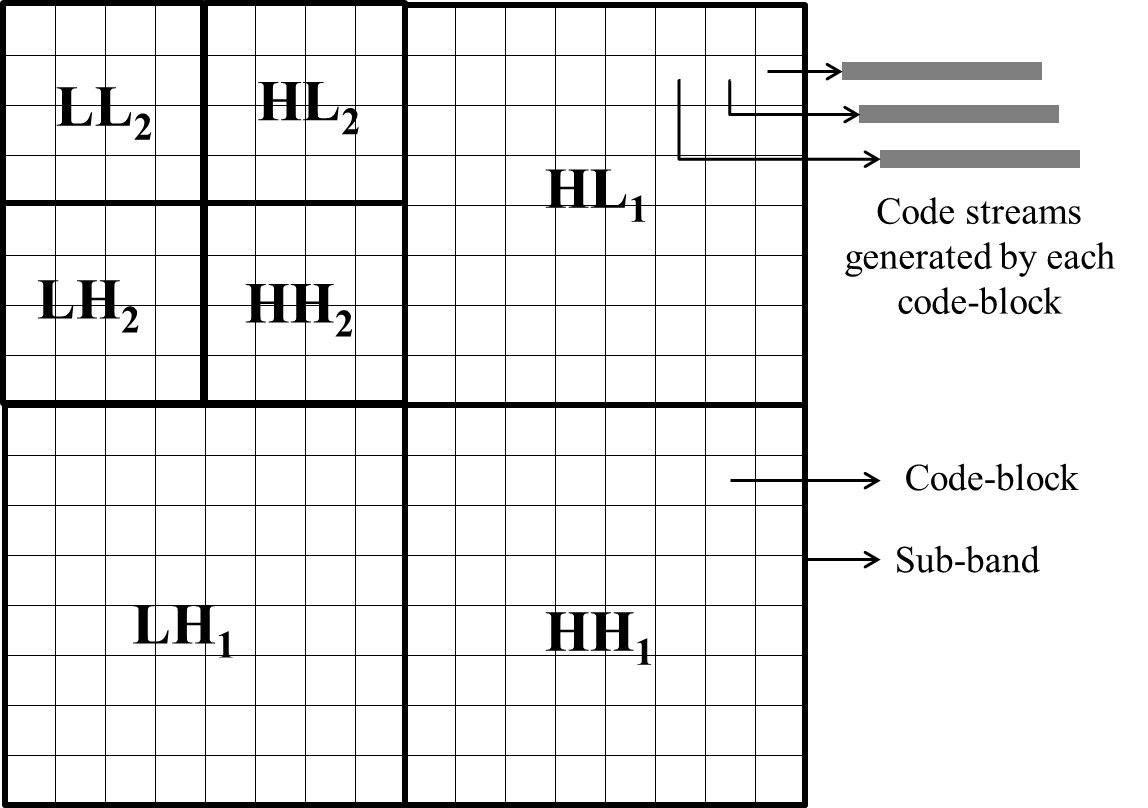}
 \caption{Illustration of Code-blocks and Sub-bands in JPEG-2000}
 \label{fig:codeblock}
\end{figure}

In JPEG-2000, quality scalability is achieved by arranging the obtained bit streams in the form of layers as shown in Figure \ref{fig:jpeg2000_qualitylayers}  \cite{boliek2002jpeg}. To get the clear interpretation of quality layer, the basic terms code-block and sub-band are indicated in Figure \ref{fig:codeblock}. Each quality layer contains the truncation point for each code-block, thus having an interpretation of the overall image quality. As per the experimentation done for JPEG-2000, it is found that the number of quality layers should be approximately twice the number of sub-bit-planes to achieve optimal quality scalability. Increased  number of layers may create the same quality reconstructed at different rates which are approximately same, causing an increase in the overhead \cite{boliek2002jpeg}. Hence, the practice of more number of quality layers is not followed in JPEG-2000.

\subsection{Rate-distortion (R-D) optimization algorithm in JPEG-2000}
Rate (R) and Distortion (D) in JPEG-2000 should satisfy the  equations (\ref{eqn:distortion}) and (\ref{eqn:rate})  \cite{boliek2002jpeg},

\begin{equation} \label{eqn:distortion}
 D = \Sigma_i D_i^{n_i}
\end{equation}

\begin{equation} \label{eqn:rate}
 R = \Sigma_i R_i^{n_i}
\end{equation}

\noindent where, $i$ represents the current code-block number, $n_i$ is the truncation point for the code-block $B_i$. Here, the main target is to find the values of $n_i$, which minimizes \emph{D} corresponding to constrained target rate \emph{$R_{max}$} $\geq$ \emph{R}. This optimization problem can be solved by minimizing (\ref{eqn:rdmin}) which is obtained by using the well-known method of Lagrange multipliers,

\begin{equation} \label{eqn:rdmin}
 \Sigma(R_i^{n_i} - \lambda D_i^{n_i})
\end{equation}

\noindent where, $\lambda$ is the Lagrange multiplier which should be varied until the target rate is achieved with minimum distortion. To obtain the values of $\lambda$, Algorithm-\ref{algo:rdopt} need to be followed  \cite{boliek2002jpeg}. This algorithm is an iterative method to obtain the best possible values of $\lambda$ which achieves optimum distortion for a required target rate. To form  progressive quality layers, $n_i$ obtained for the calculated values of $\lambda$ are taken as the truncation points and the bits are arranged accordingly.

\begin{algorithm}[!h]
\caption{Procedure get the value of $\lambda$}
\begin{algorithmic}[1]
\REQUIRE $\lambda$
\STATE $n_i$ = 0
\STATE \{
\FOR{$k = 1, 2, 3 \dots$} 
\STATE $\varDelta R_i^k$ = $R_i^k$ $-$ $R_i^{n_i}$ 
\STATE $\varDelta D_i^k$ = $D_i^k$ $-$ $D_i^{n_i}$ 
\IF {$\frac{\varDelta D_i^k}{\varDelta R_i^k} > \lambda^{-1}$}
\STATE set $n_i$ = k 
\ENDIF
\ENDFOR
\STATE  \}

\end{algorithmic}
\label{algo:rdopt}
\end{algorithm}

% \subsection{Concerns of the quality layer in JPEG-2000}
% Quality layers achieved by the R-D optimization algorithm has some major concerns:
% \begin{enumerate}
%  \item Increased burden on encoder due to the iterative process mentioned in Algorithm-\ref{algo:rdopt}.
%  \item Lack of scalability for less or single quality layer string, i.e., if string is encoded with 4 quality layers and the receiver is not capable of streaming any one of these rates, optimal quality cannot be obtained.
% \end{enumerate}
% 
% To address these concerns, BlinQS has been proposed in this paper. It does not depend on the quality layers for Quality Scalability, instead a normal distribution based algorithm has been developed to select the code-blocks which are used for the corresponding target rate.

\section{BlinQS: Blind Quality Scalability Algorithm} \label{sec:proposed_blindquality_method}
This section introduces the proposed BlinQS image compression algorithm to achieve optimum quality at target rate without using PCRD algorithms. As discussed in section-\ref{sec:qualityscaling_JPEG2000}, R-D optimization algorithms used for generating quality layers are iterative in nature, thus they increase the computational complexity. BlinQS aims to bypass the R-D optimization algorithm and achieve near optimal quality, thus reducing the computation load on the encoder. This has been achieved by using blind selection of the code-blocks obtained through gaussian normal distribution. It has shown good approximation in chosing the code-blocks optimally for required target rate. This is because, the code-blocks are arranged in the order of descending information content and selected depending on the variance boundary in which the code-block falls. As this operation is computationally effective, it does not create any prominent load on the transcoder, where BlinQS algorithm has been placed. Thus the aim of reducing the computational complexity and memoryless scalability has been achieved using BlinQS algorithm. Complete algorithm is explained in three sections: (i) Encoding, (ii) Inclusion map, and,  (iii) Decoding. BlinQS is a part of the transcoder which forms the quality layers blindly from the encoded string using the inclusion map. Main tasks to be performed by the BlinQS trancoder on the received string are:
\begin{enumerate}
 \item Getting the value of $\delta_b$ for each sub-band.
 \item Calculating the truncation point of the code-blocks.
\end{enumerate}

\subsection{BlinQS: Encoding} \label{subsec:encoding}
\begin{figure} [!h]
\centering
\subfigure[BlinQS encoder]{\includegraphics[width=0.4\columnwidth]{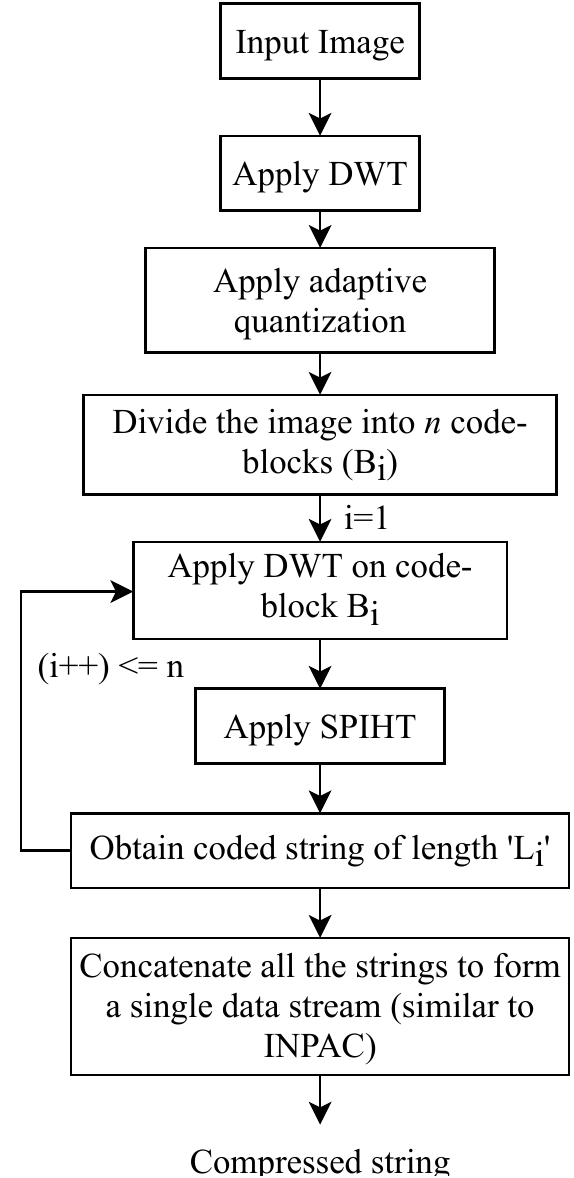}} \label{fig:blinqsencoder}
\subfigure[BlinQS decoder]{\includegraphics[width=0.48\columnwidth]{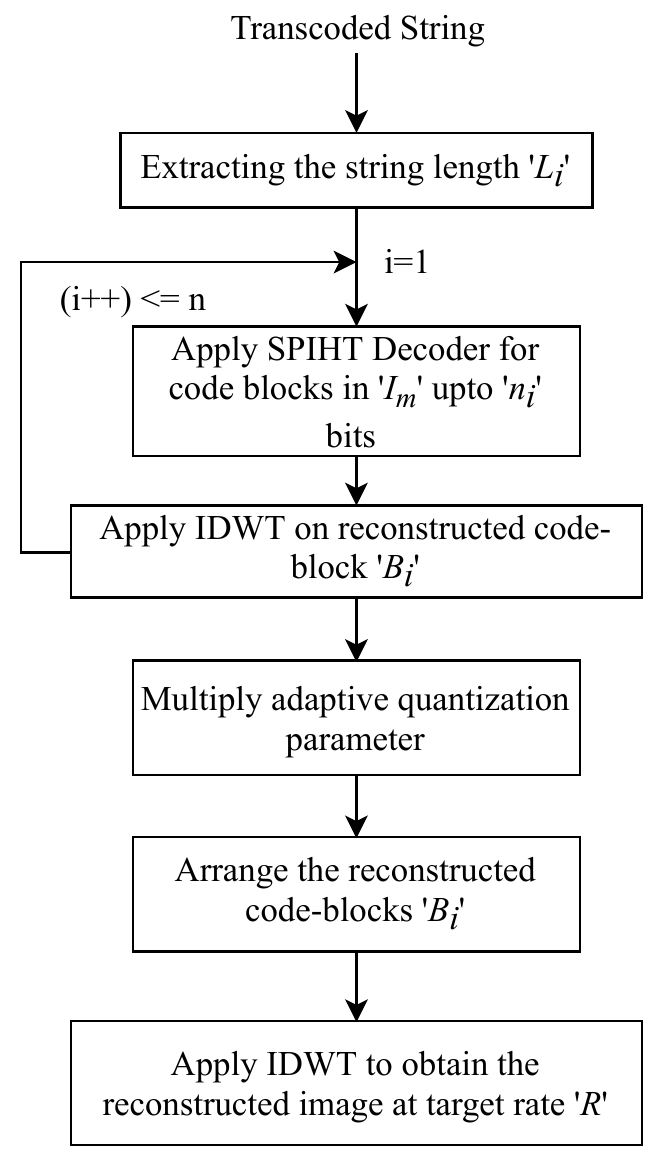}} \label{fig:blinqsdecoder} 
 
\caption{Flow Chart of Proposed BlinQS Image Compression Algorithm}
\label{fig:blinqsimagecomp}
\end{figure}

Encoding module consists of three sub-modules, (i) Image Transformation, (ii) Image Compression using Set Partition In Hierarchical Trees (SPIHT), and,  (iii) String arrangement along with the header. The encoding procedure is briefly explained in Figure \ref{fig:blinqsimagecomp}a.

\begin{figure} [!h]
\centering
\subfigure[]{\includegraphics[width=0.55\columnwidth]{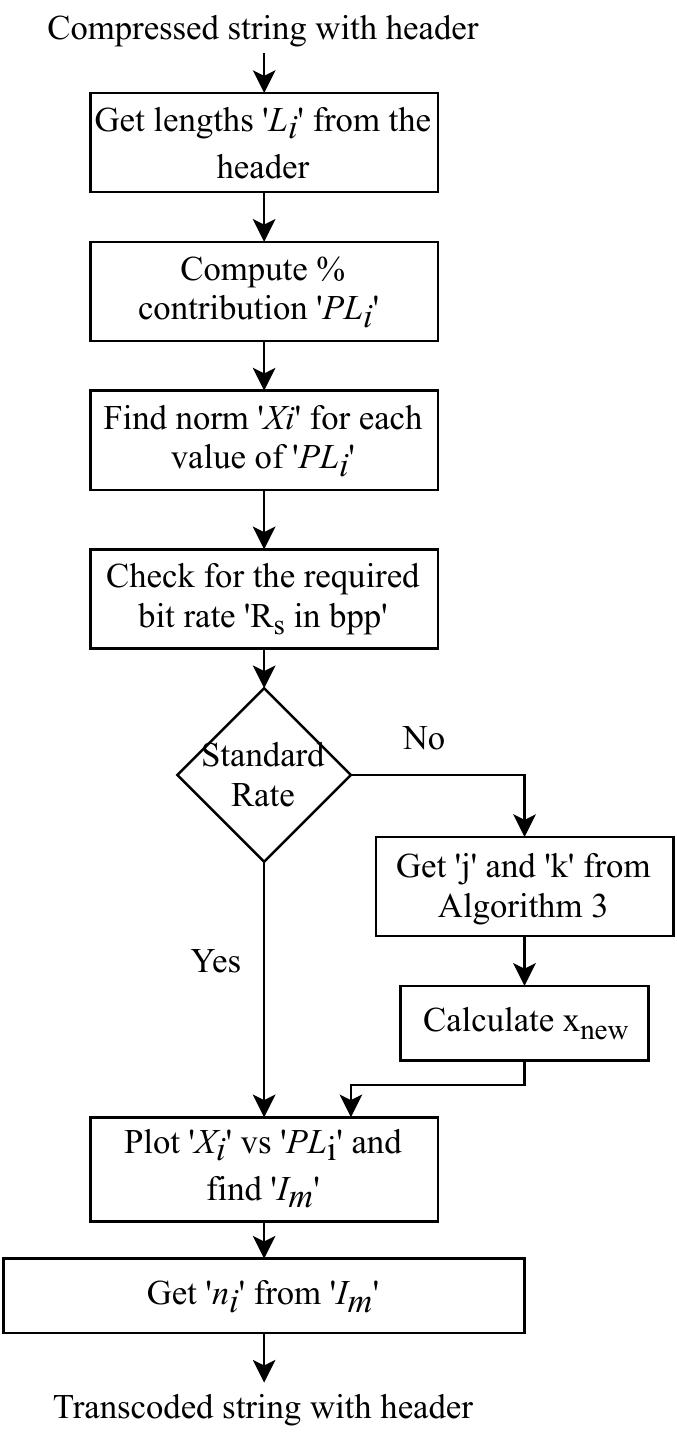}} \label{fig:blinqsinclusionmap}
 
\caption{BlinQS inclusion map selection}
\label{fig:blinqsinclmap}
\end{figure}

\subsubsection{Image Tranformation}
As shown in Figure \ref{fig:blinqsimagecomp}a, the first step of BlinQS encoding algorithm is to transform the image using Discrete Wavelet Transform (DWT).  To decompose the image, `$\emph{Biorthogonal 4.4 (bior4.4)}$' wavelet family  has been used \cite{bruni2020family}. Before decomposing, an intensity level shift of -127 is performed on the image and then it is decomposed using DWT into sub-bands.

After decomposition, the sub-bands are divided into code-blocks as illustrated in Figure \ref{fig:codeblock}, which can be considered as the building blocks of the coding. They are encoded using the SPIHT algorithm. To improve the efficiency of SPIHT, DWT is applied for each code-block in High-pass region using the same `$\emph{bior4.4}`$ wavelet family before encoding. This has shown substantial improvement in terms of compression ratio. 

\subsubsection{Image compression using SPIHT}
SPIHT is an improvement to Embedded Zerotress of Wavelet (EZW) having main characteristics of SPIHT that include: efficiency, self-adaptiveness, precise rate-control, simple and fast, and fully embedded output \cite{499834}, \cite{shapiro1993embedded}, \cite{shapiro1993smart}. SPIHT directly provides the binary output, hence, there is no need of using another algorithm for converting bits to symbols \cite{kanchi}. 

\subsubsection{Quantization factor ($\delta_b$)} \label{subsec:adaptivequant}
\begin{figure}
\centering
\includegraphics[width=1\columnwidth]{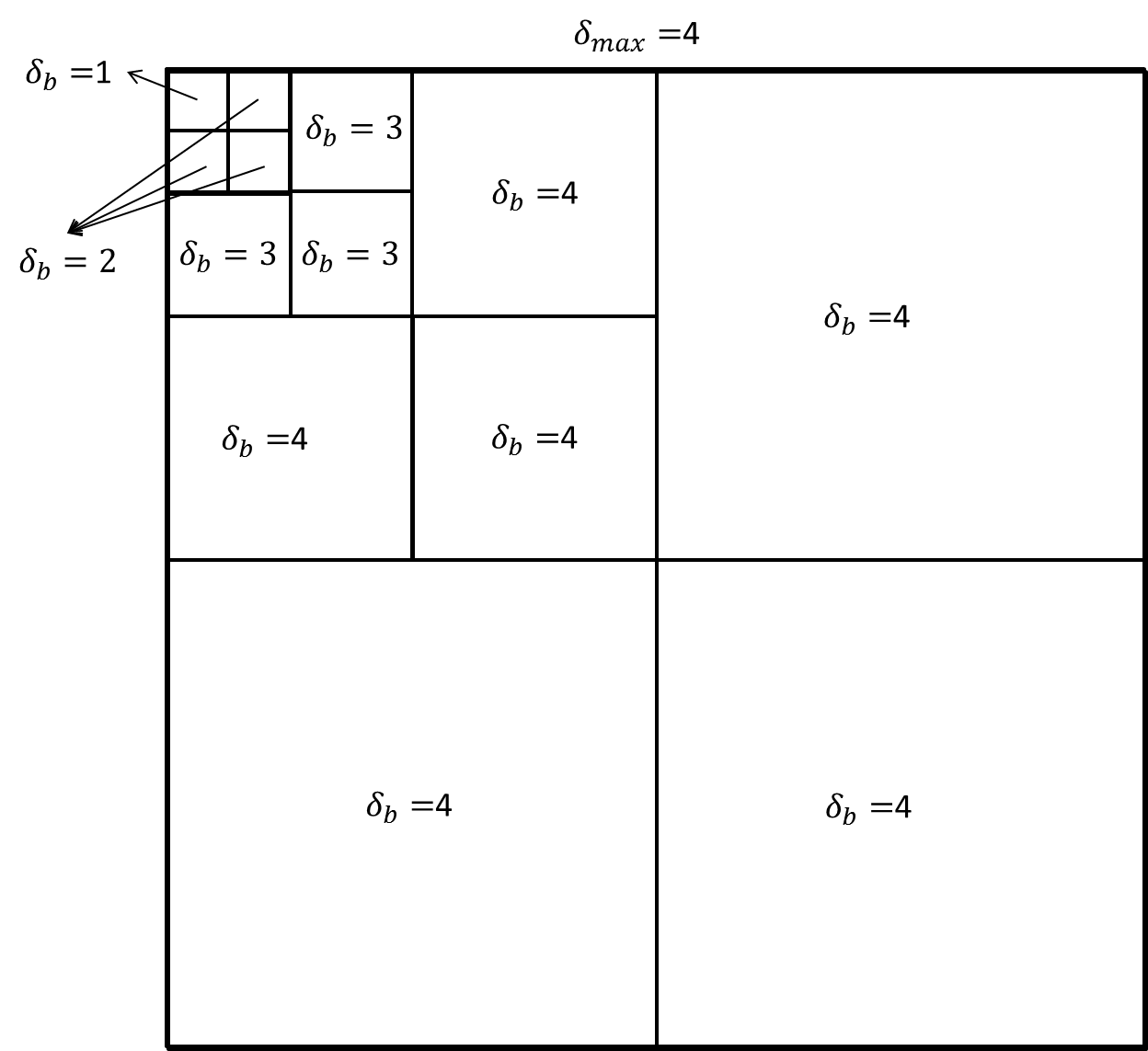}
\caption{Representation of $\delta_{adap}$ selection based on sub-band}
\label{fig:adap_delta}
\end{figure}

SPIHT algorithm is applied on each code-block (CB), after quantizing them by a factor of predefined quantization parameter ($\delta_b$) (obtained from Algorithm-\ref{algo:adapdelta}), to obtain the compressed string of the corresponding code-block. Here, quantization parameter ($\delta_b$) is the function of encoding sub-band $\delta_b = f(SB)$, i.e., the quantization factor depends on, in which sub-band the code-block is present as shown in Figure \ref{fig:adap_delta}. Let $\delta_b$ denote the quantization parameter for sub-band $SB_b$, where $b$ denotes DWT level of the sub-band. Each subband has different $\delta_b$ value, which is calculated as per Algorithm-\ref{algo:adapdelta}. Higher the value of $\delta_b$, lesser the string length ($L_i$) generated for the code-block $B_i$ of the sub-band $SB_b$ and higher the quantization error ($Q_e$) i.e, $\delta_b ~\alpha ~\frac{1}{L_i} ~\alpha~Q_e$. Therefore, to maintain the reconstruction quality of image, LL components are not quantized before coding (i.e., $\delta_b=1$) and rest of the image is quantized with $\delta_b > 1$ upto a maximum point $\delta_{max}$. Hence, $\delta_b$ is adapted with the sub-band in which the process of quantization is going on, which is termed as adaptive delta ($\delta_{adap}$).

\begin{algorithm}[!t]
\caption{Finding $\delta_b$}
\begin{algorithmic}[1]
\REQUIRE $\delta_b$
\STATE $\delta_{max}$ (Maximum value of $\delta_b$)
\STATE $\delta_b$ = 1, $SB_b$ = Maximum sub-band level ($M_{SB}$)
\STATE \{
\FOR{$b = M_{SB}, M_{SB}-1, M_{SB}-2 \dots$} 
\STATE $\delta_b$ = $\delta_b$ + 1
\IF {$\delta_b \geq \delta_{max}$}
\STATE break;
\ENDIF
\ENDFOR
\STATE  \}

\end{algorithmic}
\label{algo:adapdelta}
\end{algorithm}

Let `$L_i`$ denote the length of the compressed string obtained from the code-block `$B_i$', and `$L_{ip}$' denote the length of compressed string obtained from bit plane `$p$' of code-block `$B_i$'. For a given code-block `$B_i$', the total compressed string length `$L_i$' is the summation of the individual `$L_{ip}$' from each bit plane as mentioned in equation (\ref{eqn:compressedlength}) and complete string length `$L$' is given by equation (\ref{eqn:completestringlength}). Before transmitting or storing this string, header is formed for ease of access and flexibility in operation, details of which are given in subsection-\ref{subsubsec:headerencoder}.

\begin{equation} \label{eqn:compressedlength}
 L_i = \mathlarger{\sum}_{p=1}^{No.~of~planes} L_{ip}
\end{equation}

\begin{equation} \label{eqn:completestringlength}
 L = \mathlarger{\sum}_{i=1}^{No.~of~CB^{s}} L_i
\end{equation}

\noindent where, $N_p$ represents number of planes in the transformed image and $N_c$ represents the number of code-blocks.

\subsubsection{String arrangement along with header} \label{subsubsec:headerencoder}
The lengths $L_{ip}$, obtained from the SPIHT encoder are arranged in the order of their subbands i.e., $LL_2$, $HL_2$, $LH_2$ and so on. While arranging the string in header, strings obtained from each bit plane `$p$' of code-block `$B_i$' is considered as a separate entity and is placed in the header along with marker bytes which stores the length of the string i.e., $L_{ip}$. Therefore, effective length of the string stored is `Bits occupied by string + marker bytes'. Along with this information, basic information such as size of image, level of DWT applied on image, code-block size etc, are appended to the header, which forms the basic information header. Using the basic information header, marker byte lengths are extracted first at the transcoder to form the quality layers before decoding, which is discussed in brief in sub-section \ref{subsec:blinqs}.

\subsection{BlinQS: Inclusion map} \label{subsec:blinqs}
Blind Quality Scalability refers to \emph{``obtaining the inclusion map and truncation points for the optimum reconstruction of an image at a specified bit rate.''}  An overview of this algorithm is given in Figure \ref{fig:blinqsinclmap}. Inclusion map $(I_m)$ is a matrix which consists of the information regarding code-blocks needed for decoding the image for the specified rate. The procedure for obtaining the inclusion map generation is briefly discussed in the following steps.

\subsubsection{Calculation of string contribution: Step-1}
Calculate the precentage contribution $(PL_i)$ of each compressed code-block (CB) using equation (\ref{eqn:percentcontri}), where i=1,2,$\dots$,$N_c$.

\begin{equation} \label{eqn:percentcontri}
 PL_i = \left(\frac{L_i}{L}\right) \ast 100
\end{equation}

\subsubsection{Calculation of Normal Distribution Coefficients: Step-2}
Find the normal distribution coefficients ($X$) of array `$PL$' using the mean ($\mu$) and variance ($\sigma^2$) of the elements in the array using the equations (\ref{eqn:mean}), (\ref{eqn:variance}) and (\ref{eqn:norm}).

\begin{equation} \label{eqn:mean}
 \mu = \frac{\mathlarger{\sum} PL_i}{N_c}
\end{equation}

\begin{equation} \label{eqn:variance}
 \sigma^2 = \frac{\mathlarger{\sum}_i (PL_i - \mu)^2}{N_c}
\end{equation}

\begin{equation} \label{eqn:norm}
 X_i = f(PL_i | \mu, \sigma^2) = \frac{1}{\sqrt{2\pi\sigma^2}}~e^{\frac{-(PL_i - \mu)^2}{2\sigma^2}}
\end{equation}

% \begin{wrapfigure}{l}{0.5\columnwidth} \ContinuedFloat
% \centering
% \subfigure[BlinQS decoder]{\includegraphics[width=0.48\columnwidth]{figures/BlinQS_Decoder.pdf}} \label{fig:blinqsdecoder} 
%  
% \caption{Flow Chart of Proposed BlinQS Image Compression Algorithm}
% \label{fig:blinqsalgofig}
% \end{wrapfigure}

\begin{figure*}
 \centering
 \includegraphics[width=2\columnwidth]{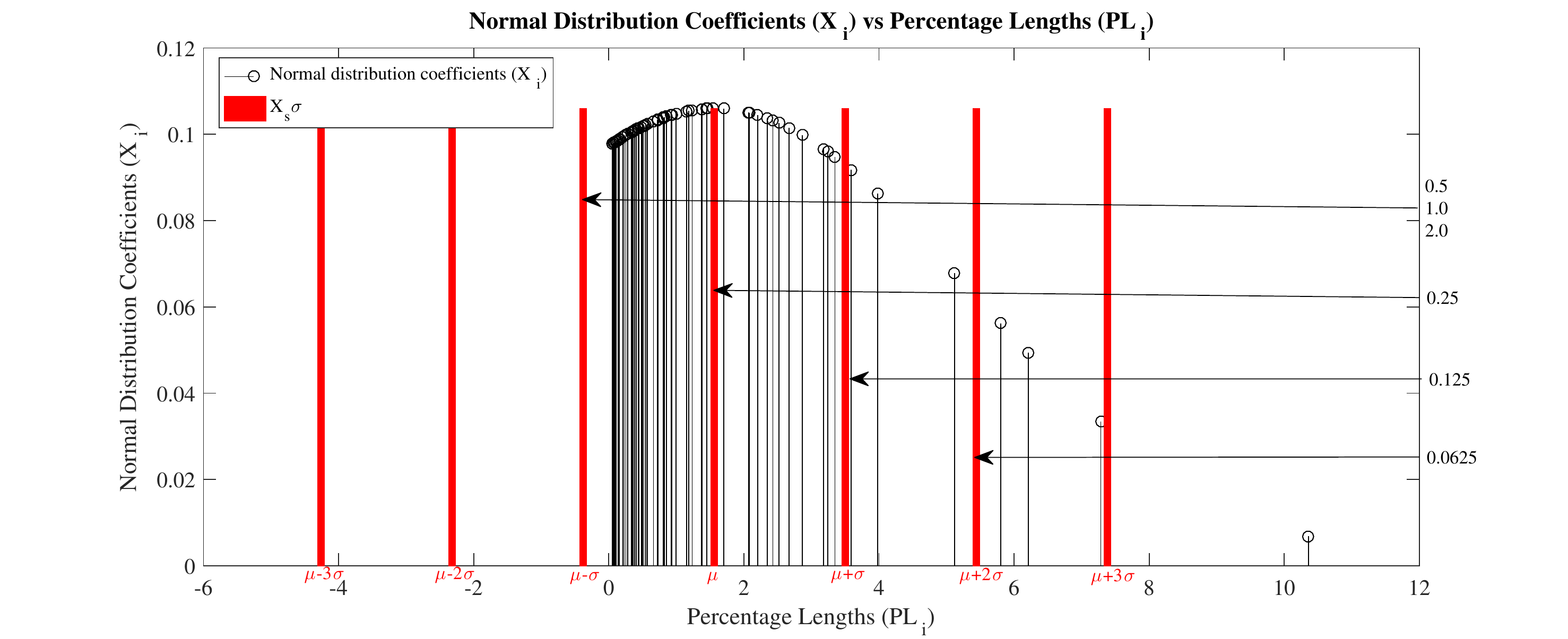}
 \caption{Normal Distribution of Percentage Lengths. Code-blocks to the right of the arrow head are considered for reconstruction for the rate (bpp) given at the arrow tail}
 \label{fig:normplot}
\end{figure*}

\subsubsection{Obtaining the code-blocks: Step-3}
Plot the values of `$X_i$', against the percentage lengths ($PL_i$) obtained from equation (\ref{eqn:percentcontri}), i.e. $X_i$ vs $PL_i$ and divide the plot by taking the step-size of $x_s\sigma$. This plot for `$Lena$' image is shown in Figure \ref{fig:normplot}. Here, the bit rates mentioned in set $R_{std} = \{0.0625, 0.125, 0.25, 0.5, 1.0, 2.0\}$ are considered as standard bit rates for which $x_s=1$, where s $\epsilon$ [1,5], and for bit rates $R_{s} < R_{new} < R_{s+1}$, a new term $x_{new}$ is introduced, which can be found out using Algorithm-\ref{algo:fractionalmultiple}. $x_{new} \epsilon (0,1)$ acts as the local step-size generator between $R_{s}$ and $R_{s+1}$ to precisely choose the code-blocks $B_i$ for the new rate $R_{new}$. Standard rates are addressed here as $R_1, R_2, \dots, R_6$ and the partition number for each rate is denoted by $loc_{R_s}$. From normal distribution plot in Figure \ref{fig:normplot}, inclusion map for standard rates can be obtained as indicated by the arrows i.e., for 0.0625: code-blocks upto $\mu~+~2\sigma$ (i.e. $loc_{R_1} = 2$), for 0.125: code-blocks upto $\mu~+~1\sigma$ (i.e. $loc_{R_2} = 3$) and so on. $x_{new}$ plays a major role in obtaining the inclusion map for $R_{new}$. If $x_{new} \epsilon$ (0,1), total number of partitions increase by a factor of 'k'. Therefore, value of $loc_{R_{new}}$ can be obtained from equation (\ref{eqn:partitionvalue}), where, $s~and~j$ are indicated in Algorithm-\ref{algo:fractionalmultiple}.

% \begin{equation} \label{eqn:partitionvalue}
%  loc_{R_{new}} = loc_{R_s}+j
% \end{equation}

\begin{subequations} \label{eqn:partitionvalue}
\begin{equation} 
 loc_{R_{new}} = loc_{R_s}+j ~ if \Delta_L < \Delta_H 
\end{equation}
\begin{equation} 
 loc_{R_{new}} = loc_{R_s}+(k-j) ~ if \Delta_L > \Delta_H 
\end{equation}
\end{subequations}

\begin{algorithm}[!t]
\caption{Finding $x_{new}$}
\begin{algorithmic}[1]
\REQUIRE $x_{new}=\frac{j}{k+1}$, where, $j, k~\epsilon$ Z
\STATE known values, $R_{std}$, $R_{new}$, $k_{max}$
\STATE find the location of $R_{new}$
\STATE $R_s < R_{new} < R_{s+1}$, $s~\epsilon$ [1,5]
\STATE get $\Delta_L$ = $|R_s - R_{new}|$ and $\Delta_H$ = $|R_{s+1} - R_{new}|$
% \IF {$\Delta_L$ $<$ $\Delta_H$}
\STATE assign $k=1$
\WHILE {$k \leq k_{max}$}
\STATE assign $j=1$
\WHILE {$j \leq k$}
\IF {$\beta_j$ $>$ T}
\IF {$\Delta_L < \Delta_H$}
\STATE $R_{new_k}$ = $R_s + (R_{s+1} - R_s) \ast \frac{j}{k+1}$
\ENDIF
\IF {$\Delta_L > \Delta_H$}
\STATE $R_{new_k}$ = $R_{s+1} - (R_{s+1} - R_s) \ast \frac{j}{k+1}$
\ENDIF
\STATE $\beta_j$ = $|R_{new} - R_{new_k}|$
\STATE $j++$
\ELSE
\STATE break;
\ENDIF
\ENDWHILE
\STATE $k++$
\ENDWHILE
% \ENDIF
\STATE obtain $j$ and $k$

% \IF {$\Delta_L$ $>$ $\Delta_H$}
% \STATE assign $j=1 and k=1$
% \WHILE {$k \leq k_{max}$}
% \WHILE {$j \leq k$}
% \IF {$\beta_j$ $>$ T}
% \STATE $R_{new_k}$ = $R_{s+1} - (R_{s+1} - R_s) \ast \frac{j}{k+1}$
% \STATE $\beta_j$ = $|R_{new} - R_{new_k}|$
% \STATE $j++$
% \ENDIF
% \ENDWHILE
% \STATE $k++$
% \ENDWHILE
% \ENDIF
% \STATE obtain $j$ and $k$

\IF {$\Delta_L$ == $\Delta_H$}
\STATE assign $k=1$ and $j=1$
\ENDIF

\end{algorithmic}
\label{algo:fractionalmultiple}
\end{algorithm}

\subsubsection{Calculation of inclusion map: Step-4}
After obtaining $loc_{R_{new}}$ and $x_{new}$ values, all the code-block strings present ahead of $loc_{R_{new}}$ are included in the inclusion map $(I_m)$. The included code-blocks can be determined from equation (\ref{eqn:inclusionmap}) as indicated below.

\begin{subequations} \label{eqn:inclusionmap}
\begin{equation} 
 I_m = CB_i ~\forall f^{-1}(X_i|\mu, \sigma^2) \geq (x_s + x_{new})\sigma ~ if \Delta_L < \Delta_H 
\end{equation}
\begin{equation} 
 I_m = CB_i ~\forall f^{-1}(X_i|\mu, \sigma^2) \geq (x_{s+1} - x_{new})\sigma ~ if \Delta_L > \Delta_H 
\end{equation}

\end{subequations}

\subsubsection{Calculation of truncation points: Step-5}
The truncation points are obtained by solving equation (\ref{eqn:blinqsrate}), which considers the code-blocks in the inclusion map ($I_m$).

After obtaining the inclusion map, truncation points `$n_i$' for each code-block `$B_i$' in the inclusion map have to be identified for a specified target rate `$R_{max}$'. The truncation points obtained here must follow the conditions specified in equations (\ref{eqn:rate}) and (\ref{eqn:blinqsrate}). On solving equation (\ref{eqn:blinqsrate}), $n_i\approx R/\Sigma_iL_i$,
where, $i$ follows the values obtained from the inclusion map. On obtaining the values of $n_i$ for each code-block in the inclusion map, a new header with the obtained lengths and string is formed and sent to decoder for image decoding.

\subsection{BlinQS: Decoding} \label{subsec:decoding}
Firstly, in decoding module, bits received are separated and required information is extracted from the header formed in section-\ref{subsec:blinqs}. Functionalities of the decoder are summarized in the flowchart given in Figure \ref{fig:blinqsinclmap}. Value of $\delta_b$ for each sub-band is calculated as per the algorithm given in Algorithm-\ref{algo:adapdelta}. After obtaining the adaptive delta, truncation points `$n_i$' for each code-block `$B_i$' have to be identified for a specified target rate `$R_{max}$'. The truncation points obtained here must follow the conditions specified in equation (\ref{eqn:blinqsrate}).

\begin{equation} \label{eqn:blinqsrate}
 \mathlarger{\sum}_i~ L_i * n_i  = R \leq R_{max}
\end{equation}

On solving equation (\ref{eqn:blinqsrate}), $n_i\approx R/\Sigma_iL_i$. On obtaining the values of $n_i$ for each code-block, SPIHT decoding is applied to the respective blocks up to the truncation points `$n_i$'.
Inverse SPIHT is applied on the blocks in the inclusion map upto the obtained truncation points. $\delta_b$ obtained from Algorithm-\ref{algo:adapdelta} is multiplied with the corresponding block values and then Inverse Discrete Wavelet Transform (IDWT) is applied using \emph{'bior4.4'} wavelet family and arranged in image format. The blocks which are not available in the inclusion map are filled with zeros and image is transformed into spatial domain by applying IDWT. And finally, the transformed image is level shifted by ``+127'' to get the reconstructed image.

\section{Results and Discussions} \label{sec:results}
This section presents the performance comparison of BlinQS with JPEG-2000 standard. Results have been presented at standard and non-standard rates for three (3) standard images: Lena (512 $\times$ 512), Barbara (512 $\times$ 512) and Elaine (512 $\times$ 512). This section is divided into three subsections: (i) Inclusion map and truncation points, (ii) Quantitative and qualitative comparison of BlinQS, and (iii) Computational complexity and trade-off.	

% \begin{figure*} [!t]
% \centering
%  \includegraphics[width=2.0\columnwidth]{figures/strlenvscodeblock.png}
%  \caption{Representation of inclusion map and Truncation points for Lena image}
%  \label{fig:strlenvscodeblock}
% \end{figure*}

\begin{figure*}%\ContinuedFloat*
\centering
 \subfigure[Req rate= 0.0625bpp, Obtained PSNR= 26.35]{\includegraphics[width=0.99\columnwidth]{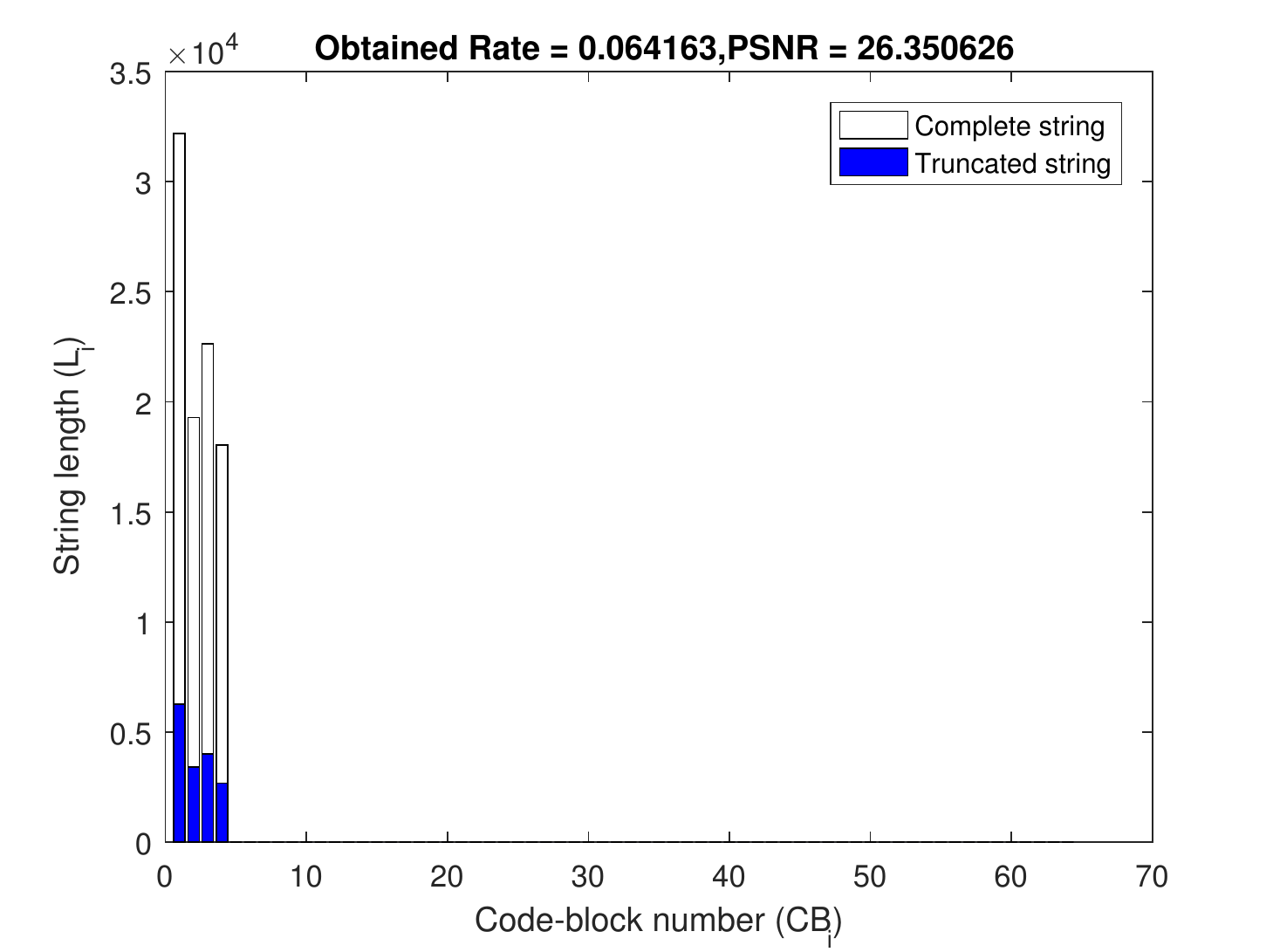}}
 \subfigure[Req rate= 0.125bpp, Obtained PSNR= 28.42]{\includegraphics[width=0.99\columnwidth]{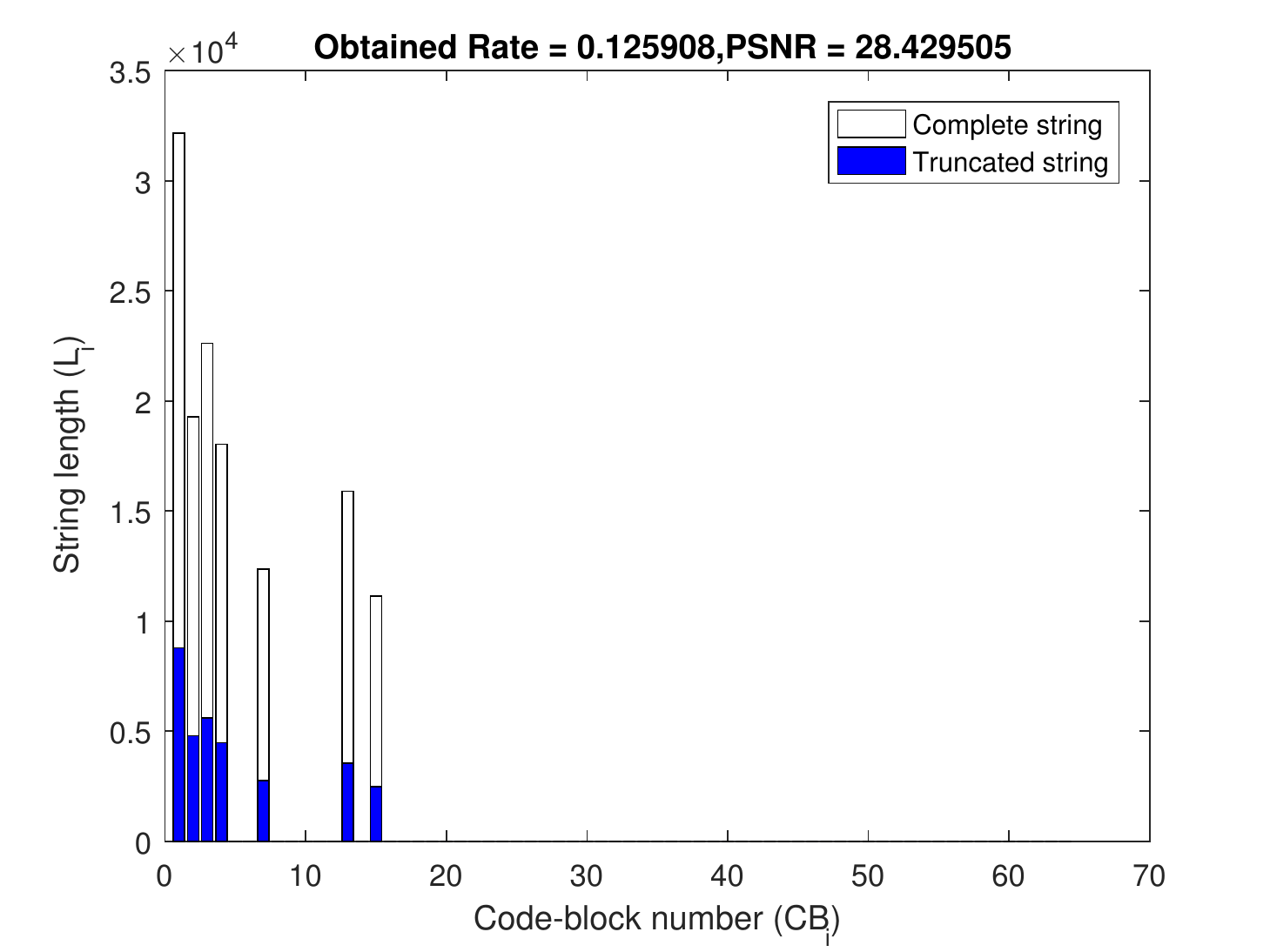}}
% \end{figure}
% \begin{figure}\ContinuedFloat
 \subfigure[Req rate= 0.25bpp, Obtained PSNR= 30.69]{\includegraphics[width=0.99\columnwidth]{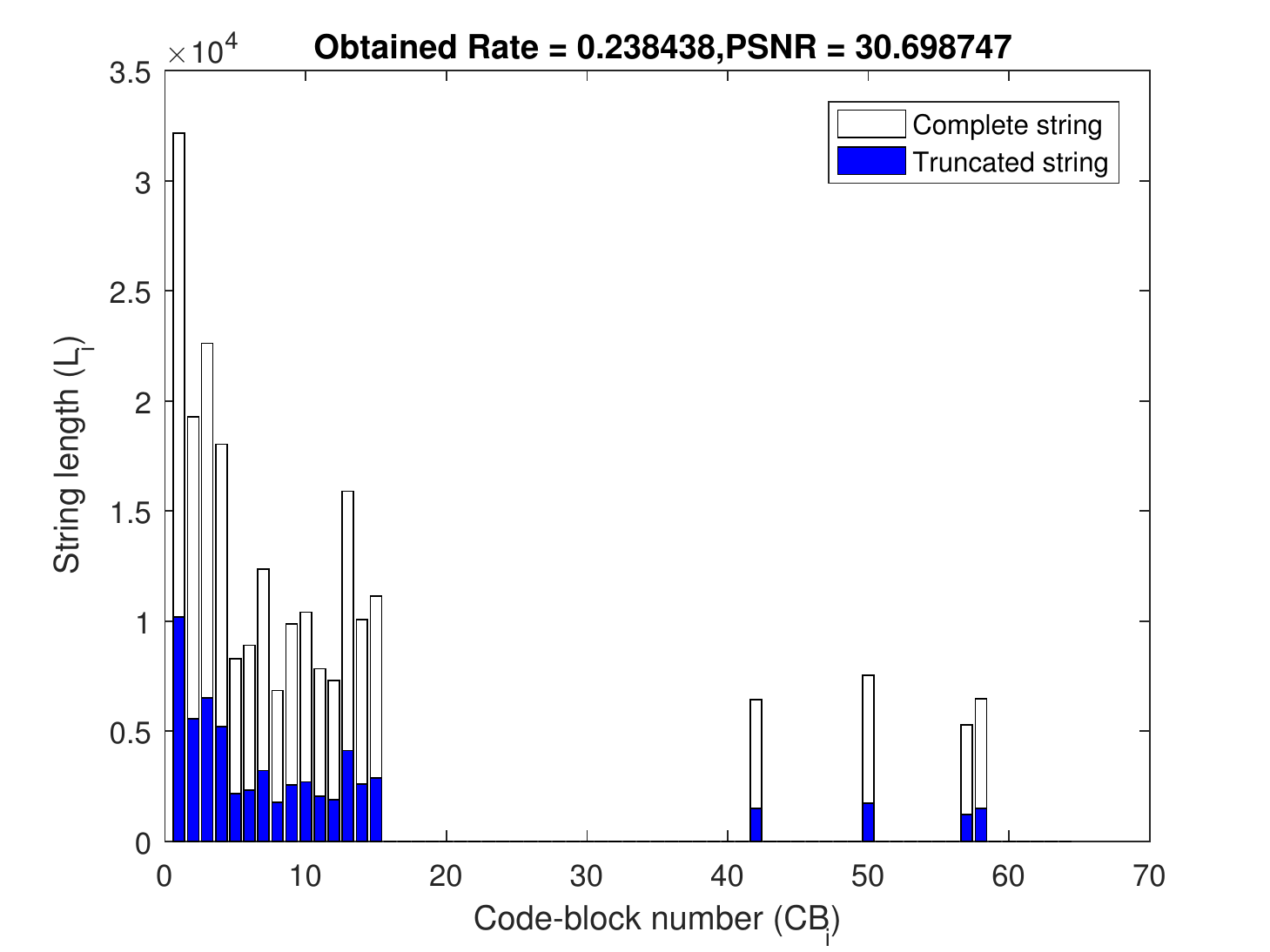}}
 \subfigure[Req rate= 0.5bpp, Obtained PSNR= 34.54]{\includegraphics[width=0.99\columnwidth]{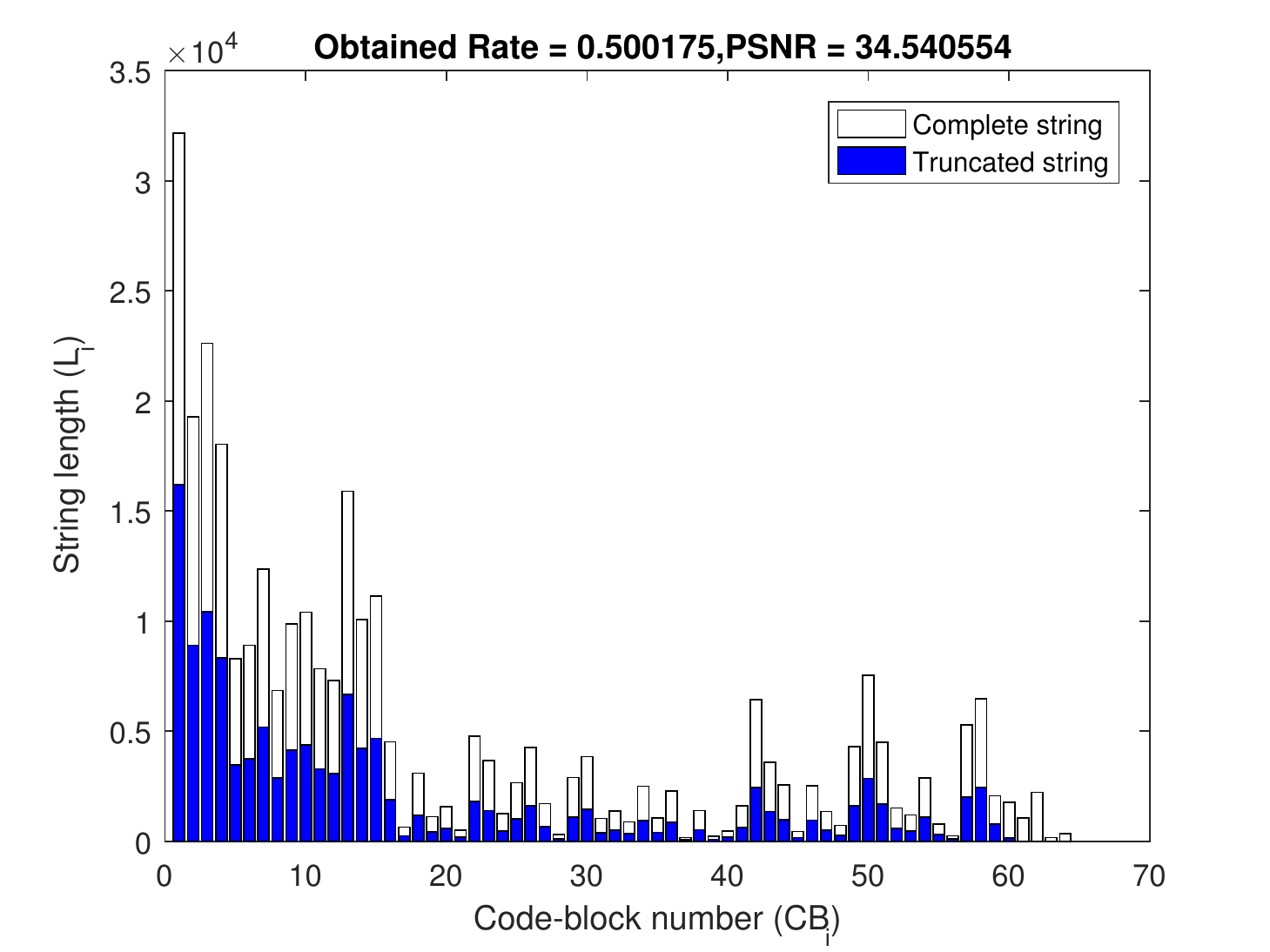}}
% \end{figure}
% \begin{figure}\ContinuedFloat
\subfigure[Req rate= 1.0bpp, Obtained PSNR= 38.22]{\includegraphics[width=0.99\columnwidth]{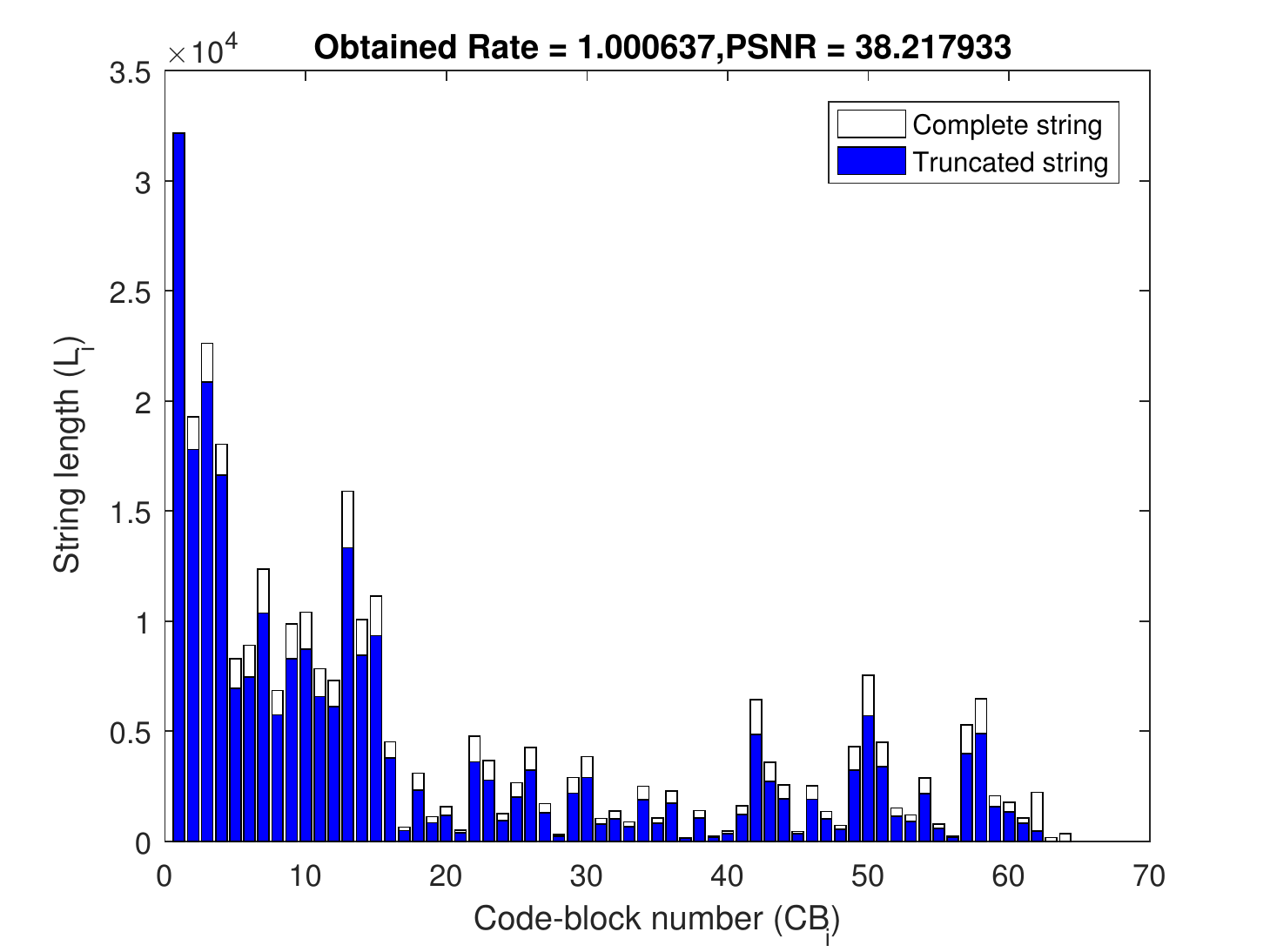}}
 \caption{Representation of inclusion map and Truncation points for Lena image}
 \label{fig:strlenvscodeblock}
\end{figure*}

\begin{figure} [!t]
\centering
 \subfigure[Lena]{\includegraphics[width=0.95\columnwidth]{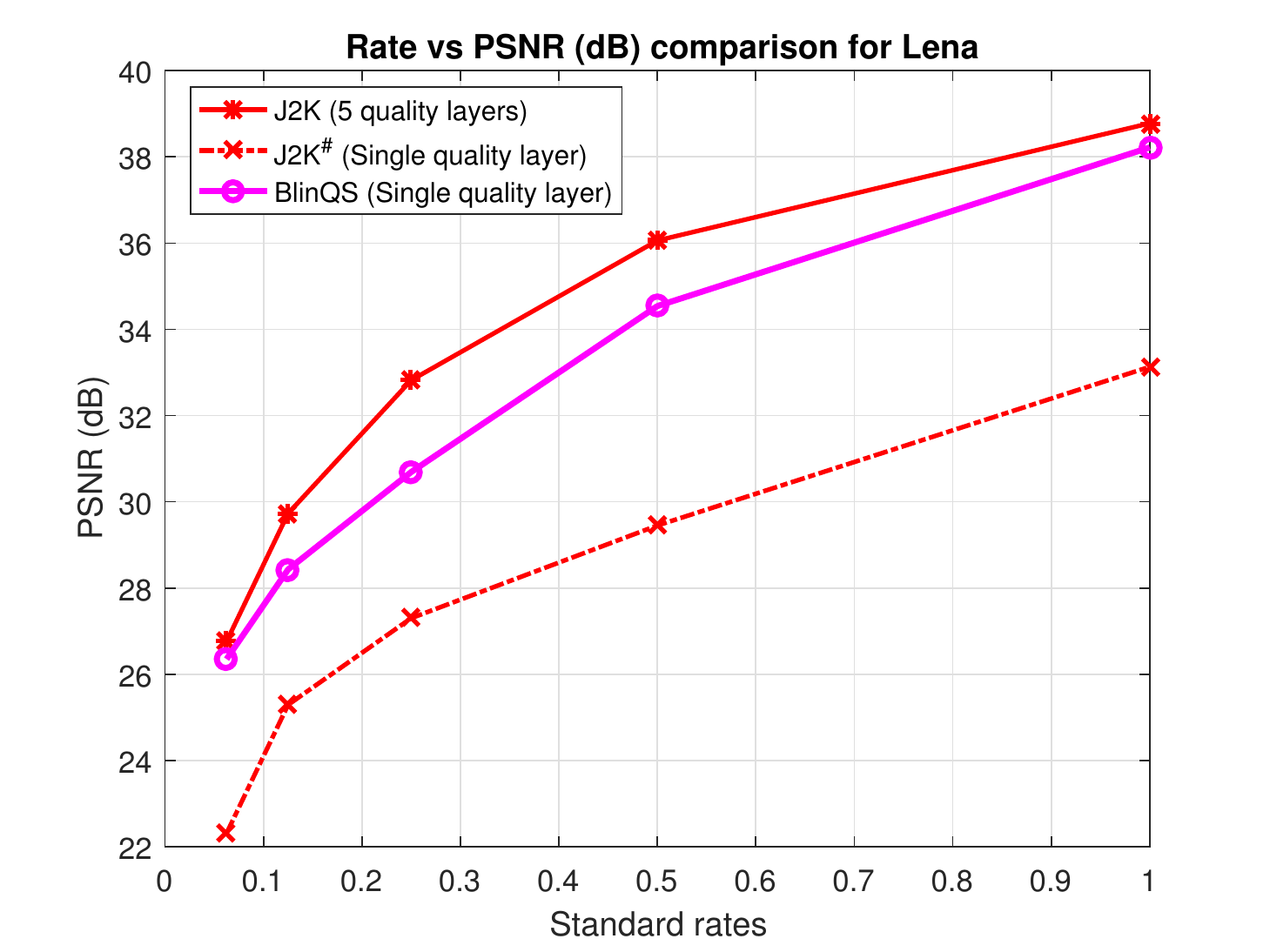}}
 \subfigure[Barbara]{\includegraphics[width=0.95\columnwidth]{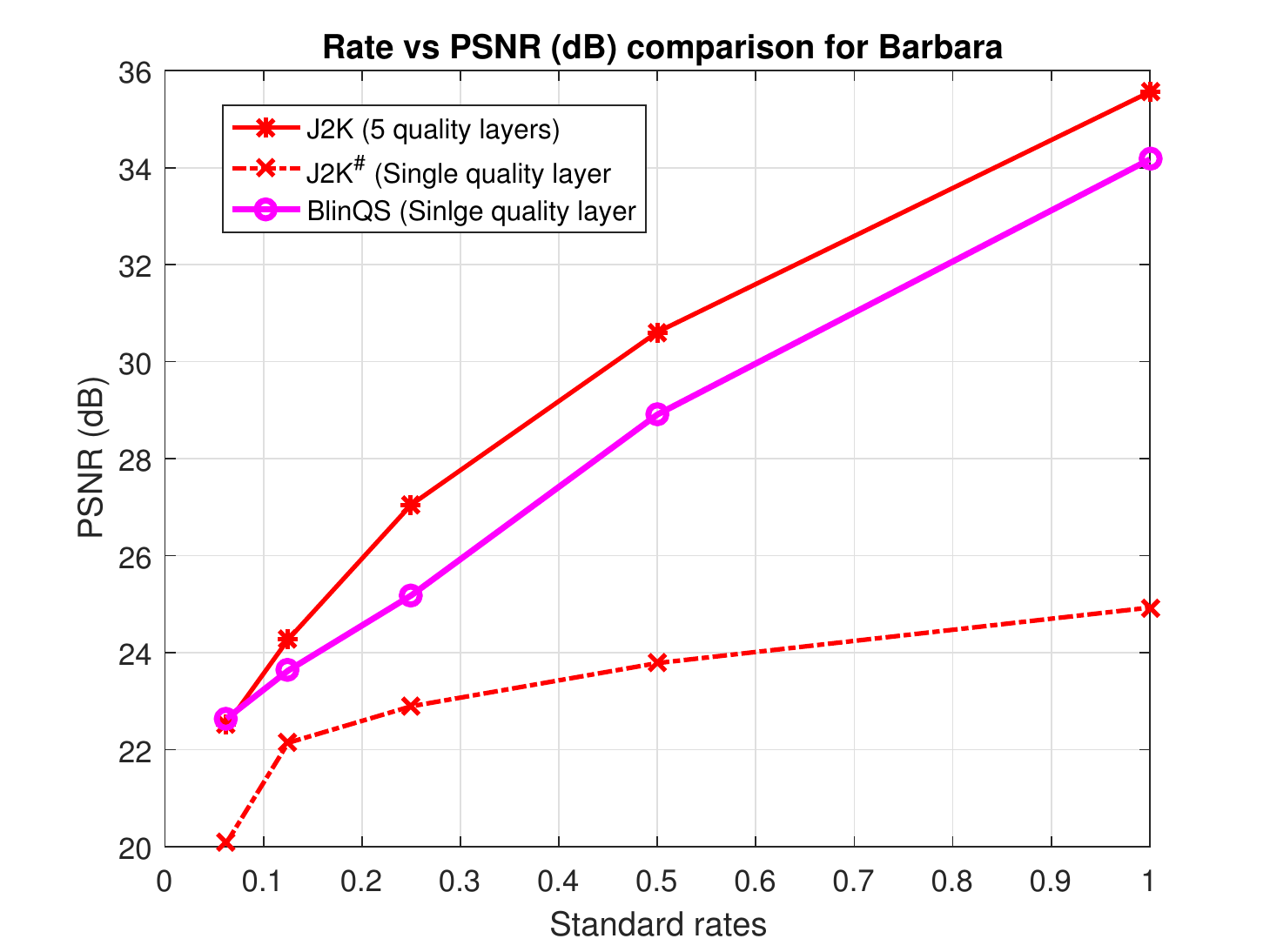}}
 \caption{Rate vs PSNR (dB)}
 \label{fig:psnrvsrate}
\end{figure}

\subsection{Inclusion map and the truncation points}
The inclusion map and the truncation points of the corresponding code-blocks for Lena image are presented in Figure \ref{fig:strlenvscodeblock}. In this figure, the white bar indicates the complete length of the string $(L_i)$ obtained for code-blocks $B_i$ and the black bar indicates the truncation point for $B_i$ i.e., amount of string used for reconstruction for the target rate. Obtained Peak Signal to Noise Ratio (PSNR) values for the standard rates mentioned in $R_{std}$ are shown in Figure \ref{fig:strlenvscodeblock}. These values clearly show the effect of inclusion map and the truncation points obtained using BlinQS. For required rates, 0.5 and 1.0, the inclusion map obtained is same as shown in Figure \ref{fig:normplot}, but the truncation points for these rates are different. Therefore, it is clear that the truncation points have played a major role in providing good quality at that rates. To optimally maintain the quality, BlinQS does not pick the blocks in the order, instead it picks up the blocks in the order of their contribution to the quality which can be derived using $X_i$. This can be clearly seen for the target rates 0.125 and 0.25 in Figure \ref{fig:strlenvscodeblock}, where some of the code-blocks are skipped by the algorithm to achieve optimum quality. For obtaining the optimum quality for the non-standard rates $R_{new}$, the local step size $x_{new}$ plays a vital role in obtaining the inclusion map $I_m$, which is obtained from Algorithm-\ref{algo:fractionalmultiple}.

% Table generated by Excel2LaTeX from sheet 'For paper'
\begin{table}[htbp]
  \centering
  \caption{PSNR (dB) comparison of BlinQS and JPEG-2000 at standard rates}
    \label{table:compstdrates} 
    {\renewcommand{\arraystretch}{1.5}
    \begin{tabular}{clrrr}
    \hline
    \multicolumn{1}{l}{\textbf{Image}} & \textbf{Rate (bpp)} & \multicolumn{1}{c}{\textbf{J2K}} & \multicolumn{1}{c}{\textbf{J2K*}} & \multicolumn{1}{c}{\textbf{BlinQS}} \\
    \hline
    \multirow{5}[10]{*}{Lena} & 0.0625 & 26.76 & 22.31 & 26.35 \\
           & 0.125 & 29.73 & 25.29 & 28.43 \\
           & 0.25  & 32.82 & 27.3  & 30.69 \\
           & 0.5   & 36.06 & 29.45 & 34.54 \\
           & 1     & 38.78 & 33.13 & 38.22 \\
    \hline
    \multirow{5}[10]{*}{Barbara} & 0.0625 & 22.51 & 20.08 & 22.64 \\
           & 0.125 & 24.27 & 22.141 & 23.63 \\
           & 0.25  & 27.05 & 22.894 & 25.18 \\
           & 0.5   & 30.61 & 23.785 & 28.91 \\
           & 1     & 35.56 & 24.929 & 34.17 \\
    \hline
    \multicolumn{1}{c}{\multirow{5}[10]{*}{Elaine}} & 0.0625 & 28.17 & 23.15 & 27.61 \\
           & 0.125 & 30.35 & 27.28 & 29.39 \\
           & 0.25  & 31.79 & 28.615 & 30.51 \\
           & 0.5   & 33.03 & 30.88 & 31.82 \\
           & 1     & 35.12 & 32.66 & 34.37 \\
    \hline
    \end{tabular}%
    }
    
%     \vspace{4pt}
%     \begin{minipage}{0.4\textwidth}
      \flushright{\footnotesize{J2K: JPEG-2000 from \cite{kakadu2000}, \cite{openjpegmaster}, $^\#$-without quality layers}}
%     \end{minipage}
    
    \vspace{6pt}
    \begin{minipage}{\columnwidth}
      \footnotesize{*Proposed BlinQS has comfortably ahead of J2K$^\#$ at all the rates and performing equally well with J2K despite using the single layered string. This adds a new degree of freedom at the user end to chose any required rate independent from the encoder.}
    \end{minipage}
\end{table}%

% Table generated by Excel2LaTeX from sheet 'Sheet2'
\begin{table}[htbp]
  \centering
  \caption{Sample Database}
    \begin{tabular}{clcc}
    \noalign{\smallskip}\hline\noalign{\smallskip}
    \textbf{S. No} & \multicolumn{1}{c}{\textbf{Image Name}} & \textbf{Resolution}& \textbf{Set}\\
    \noalign{\smallskip}\hline\noalign{\smallskip}
    1     & Baboon & \multirow{4}[8]{*}{512$\times$512} & \multirow{4}[8]{*}{Set-I}\\
\noalign{\smallskip}    2     & Plane &  & \\
\noalign{\smallskip}   3     & Peppers &  & \\
\noalign{\smallskip}    4     & Ship  &  & \\
    \noalign{\smallskip}\hline\noalign{\smallskip}
    5     & Boat & \multirow{2}[4]{*}{3840$\times$2160} & \multirow{2}[4]{*}{Set-II (4K)}\\
\noalign{\smallskip}    6    & Sand  &  & \\
    \noalign{\smallskip}\hline
    \end{tabular}%
  \label{table:sampledb}%
  
  \vspace{6pt}
  \begin{minipage}{\columnwidth}
  \footnotesize{*Apart from the standard images, results for UHD and other standard images are taken for comparison. More images are taken for comparison in Appendix \ref{A1}}
  \end{minipage}
\end{table}%

% \begin{table*} [!t]
% \centering
% % table caption is above the table
% \caption{PSNR (dB) comparison of BlinQS and JPEG-2000 at standard rates}
% \label{table:compstdrates}       % Give a unique label
% % For LaTeX tables use
% % \resizebox{\columnwidth}{!}{
% % \begin{tabular}{|p{1cm}|p{1cm}|p{1cm}|p{1cm}|p{1cm}|p{1cm}|p{1cm}|p{1cm}|p{1cm}|}
% \begin{tabular}{|c|c|c|c|c|c|c|c|c|}
% \hline
% Rate & \multicolumn{2}{|c|}{Lena (512$\times$512)} & \multicolumn{2}{|c|}{Barbara (512$\times$512)} & \multicolumn{2}{|c|}{Woman (2560$\times$2048)} & \multicolumn{2}{|c|}{Bike (2560$\times$2048)} \\
% \cline{2-9}
%  & B & J \cite{taubman2000high} & B & J \cite{taubman2000high} & B & J \cite{taubman2000high} & B & J \cite{taubman2000high} \\
%  \hline
%  0.0625 & 26.35 & 28.10 & 22.64 & 23.34 & 24.01 & 25.63 & 20.26 & 23.78  \\
%  \hline
%  0.125 & 28.43 & 31.05 & 23.63 & 25.37 & 25.13 & 27.39 & 21.79 & 26.37 \\
%  \hline
%  0.25 & 30.69 & 34.16 & 25.18 & 28.40 & 26.85 & 30.04 & 24.51 & 29.60 \\
%  \hline
%  0.5 & 34.54 & 37.29 & 28.91 & 32.29 & 31.06 & 33.70 & 29.20 & 33.46 \\
%  \hline
%  1.0 & 38.22 & 40.48 & 34.17 & 37.11 & 36.46 & 38.49 & 33.78 & 38.09 \\
% \hline
% \end{tabular}
% \vspace{2pt}
% \flushright{\footnotesize{*B: BlinQS, J: JPEG-2000 from \cite{taubman2000high}}} \hspace{1.5cm}
% % \vspace{1cm}
% % \raggedright
% % }
% \end{table*}

\begin{figure*}[!t]
\centering
 \subfigure[Lena]{\includegraphics[width=0.32\columnwidth]{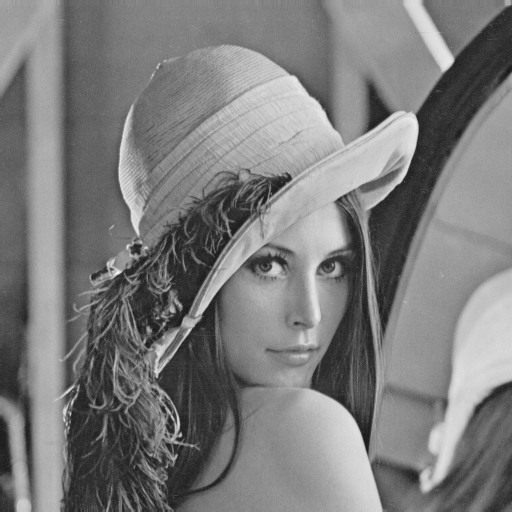}} \\
 \subfigure[Rate= 0.0625 bpp]{\includegraphics[width=0.32\columnwidth]{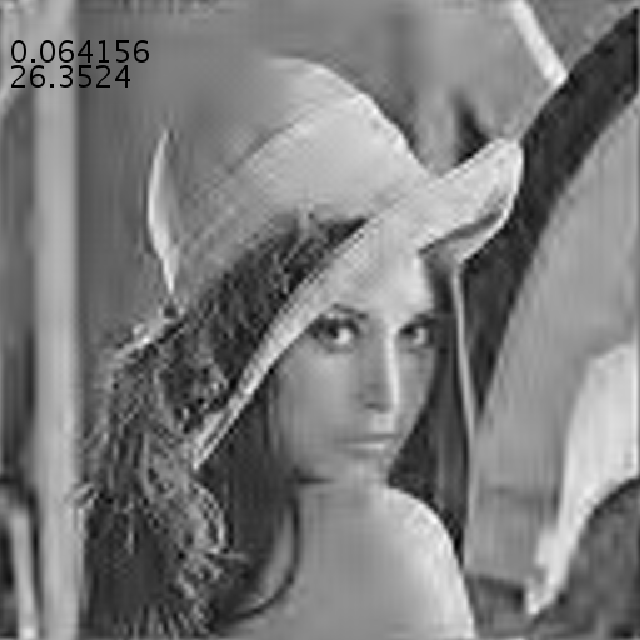}}
 \hfill
 \subfigure[Rate= 0.125 bpp]{\includegraphics[width=0.32\columnwidth]{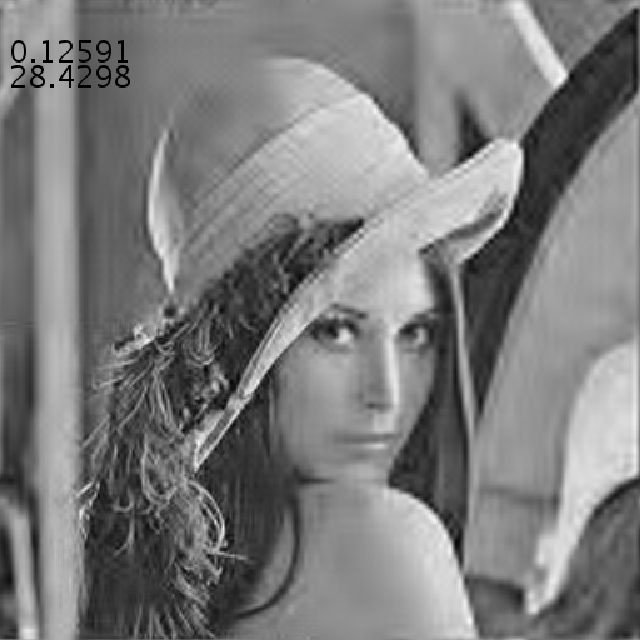}} \hfill
 \subfigure[Rate= 0.25 bpp]{\includegraphics[width=0.32\columnwidth]{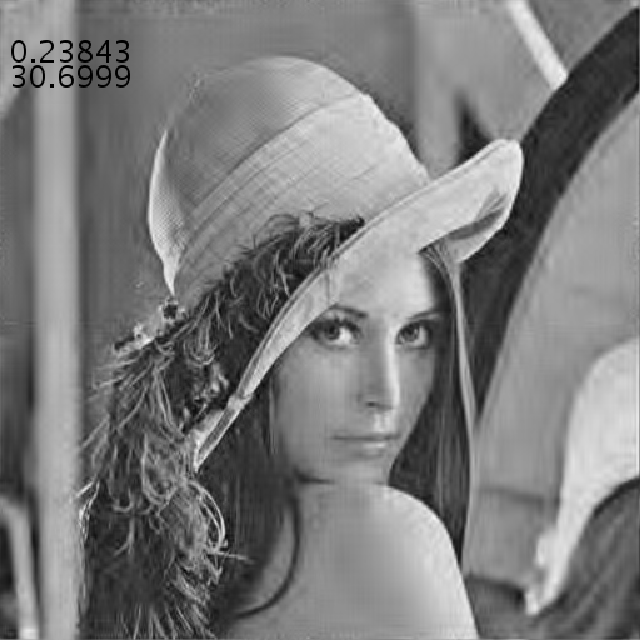}}
 \hfill
 \subfigure[Rate= 0.5 bpp]{\includegraphics[width=0.32\columnwidth]{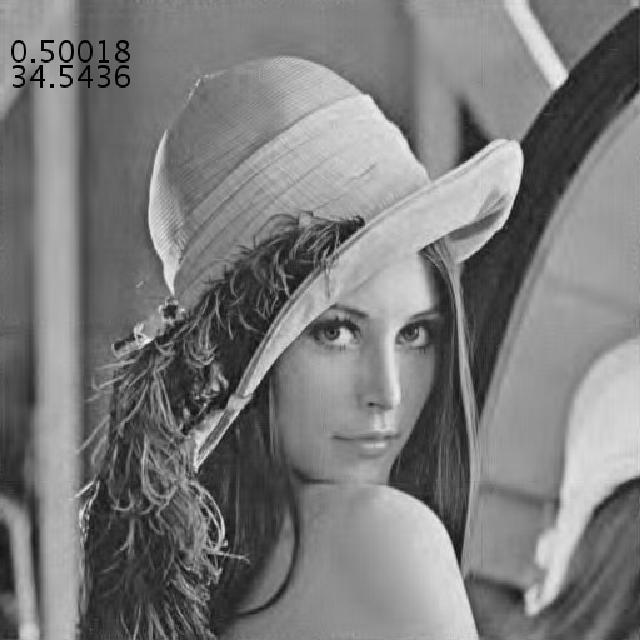}} \hfill
% \end{figure}
% \begin{figure} \ContinuedFloat
%  \centering
 \subfigure[Rate= 1.0 bpp]{\includegraphics[width=0.32\columnwidth]{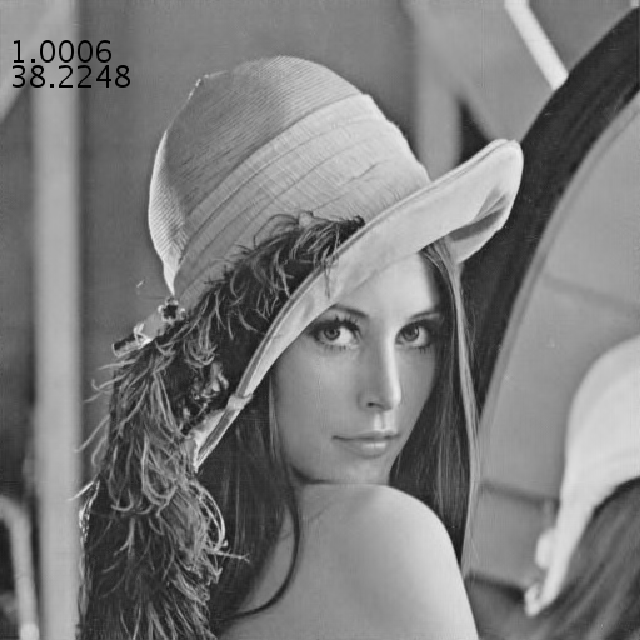}}
 \hfill
 \subfigure[Rate= 1.19]{\includegraphics[width=0.32\columnwidth]{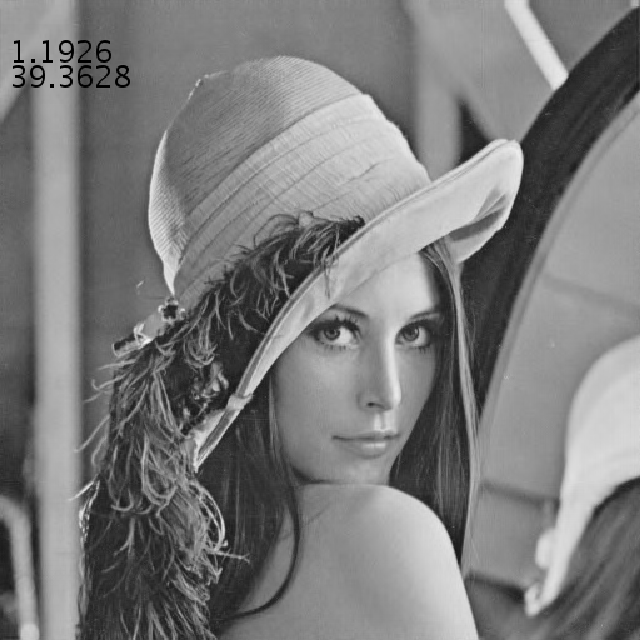}} \hfill
 \caption{Lena reconstructed at standard rates: (a) Original Image, (b)-(g) Reconstructed at specified rates using the proposed BlinQS algorithm}
 \label{fig:lena_rates}
\end{figure*}

\begin{figure*}[!t]
\centering
 \subfigure[Barbara]{\includegraphics[width=0.32\columnwidth]{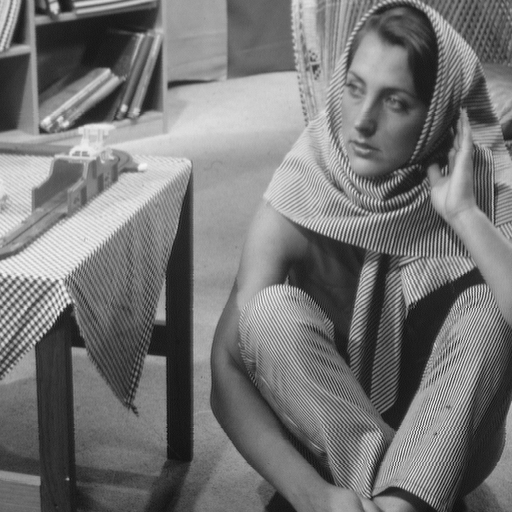}}\\
 \subfigure[Rate= 0.0625 bpp]{\includegraphics[width=0.32\columnwidth]{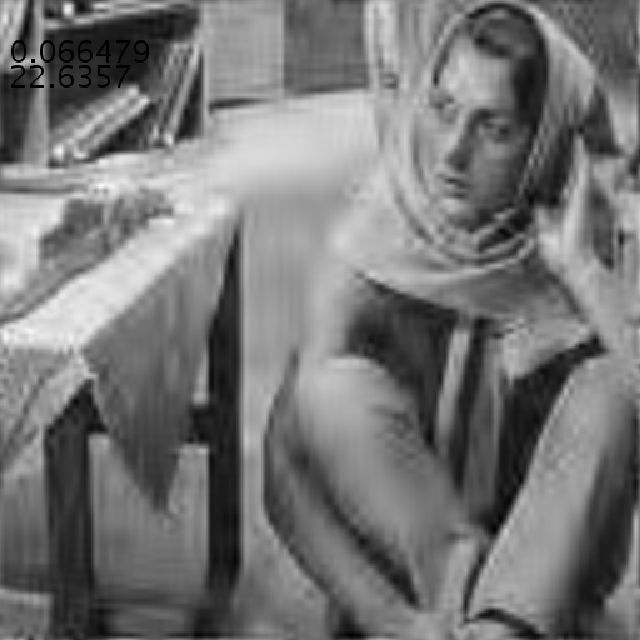}} \hfill
 \subfigure[Rate= 0.125 bpp]{\includegraphics[width=0.32\columnwidth]{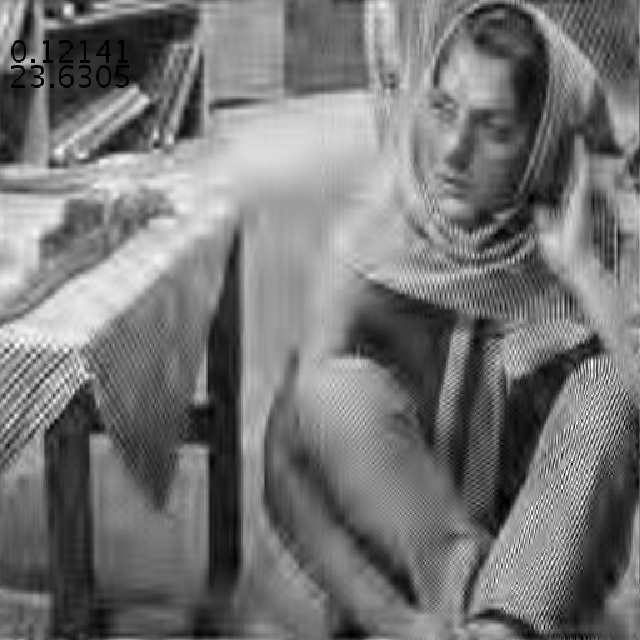}} \hfill
 \subfigure[Rate= 0.25 bpp]{\includegraphics[width=0.32\columnwidth]{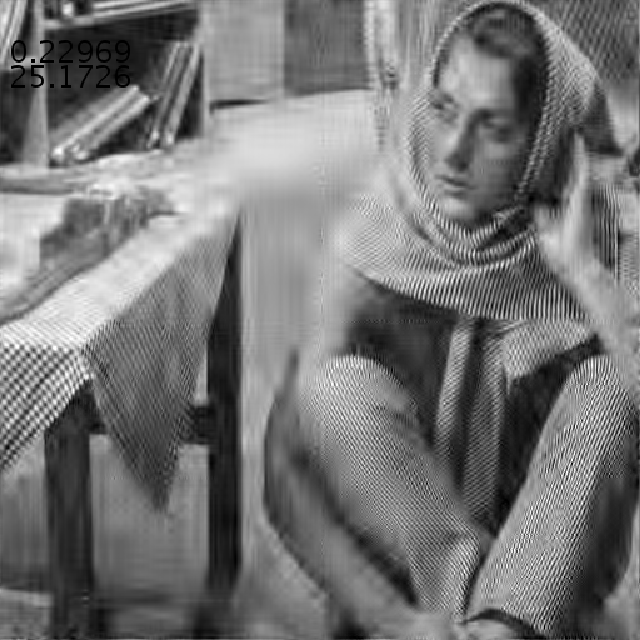}} \hfill
 \subfigure[Rate= 0.5 bpp]{\includegraphics[width=0.32\columnwidth]{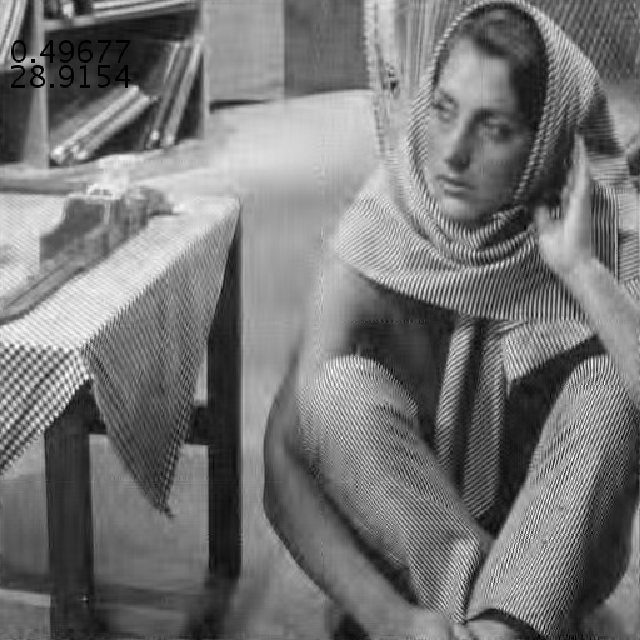}} \hfill
 \subfigure[Rate= 1.0 bpp]{\includegraphics[width=0.32\columnwidth]{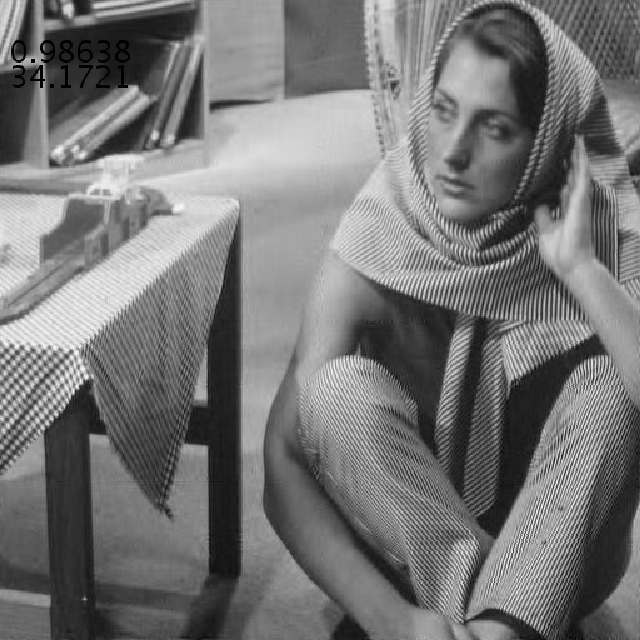}} \hfill
 \subfigure[Rate= 1.73 bpp]{\includegraphics[width=0.32\columnwidth]{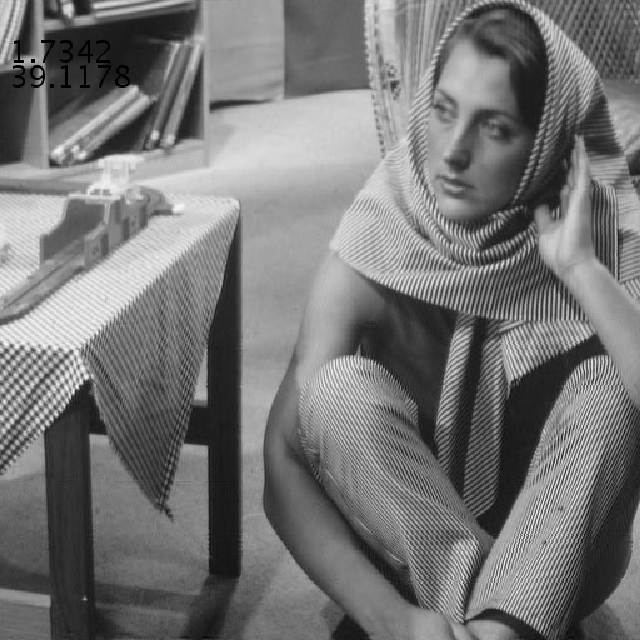}} \hfill
 \caption{Barbara reconstructed at standard rates: (a) Original Image, (b)-(g) Reconstructed at specified rates using the proposed BlinQS algorithm}
 \label{fig:barbara_rates}
\end{figure*}

\begin{figure*}[!t] 
\centering
 \subfigure[boat]{\includegraphics[width=0.32\columnwidth, height=2.5cm]{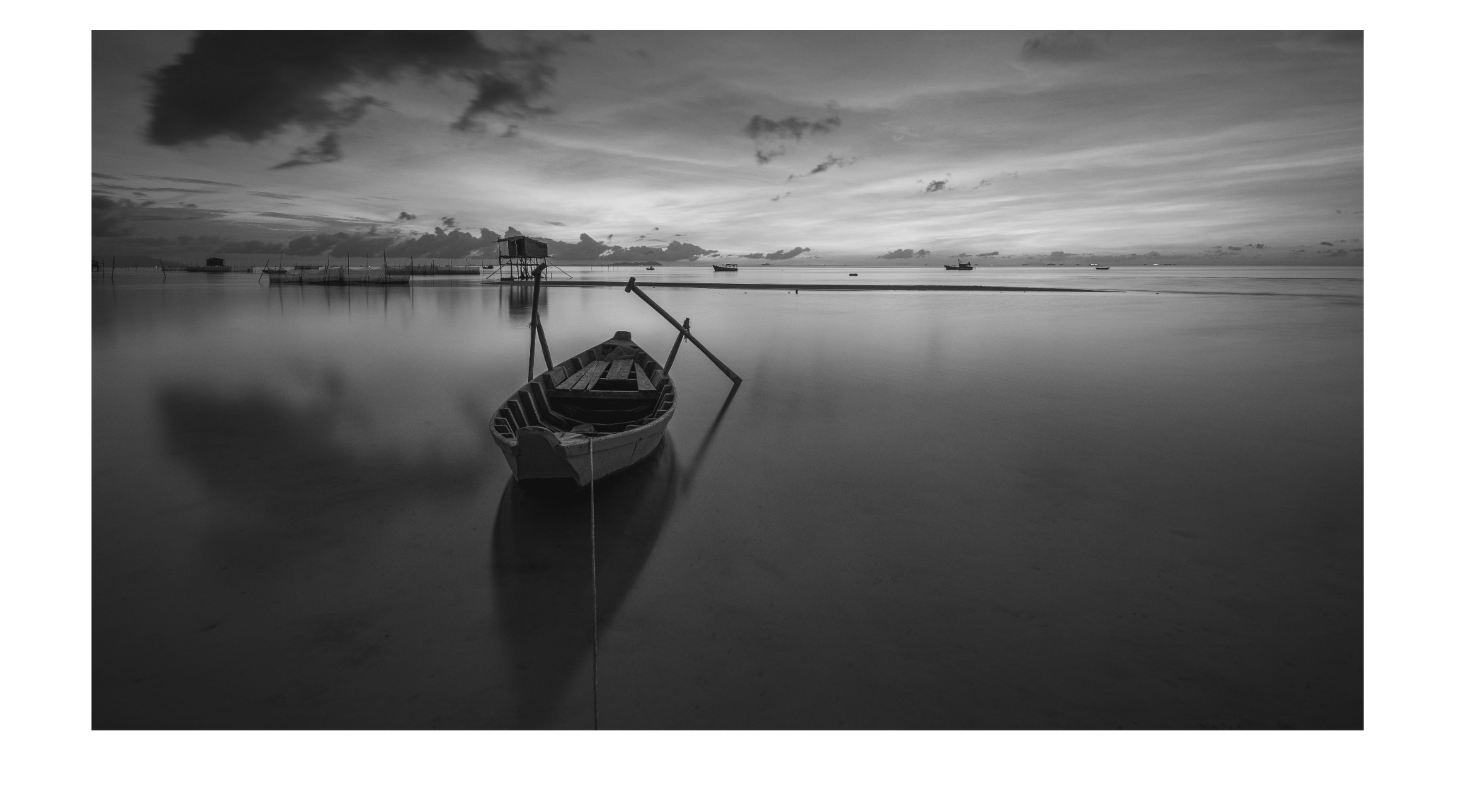}}\\
 \subfigure[Rate= 0.0625 bpp]{\includegraphics[width=0.32\columnwidth, height=2.5cm]{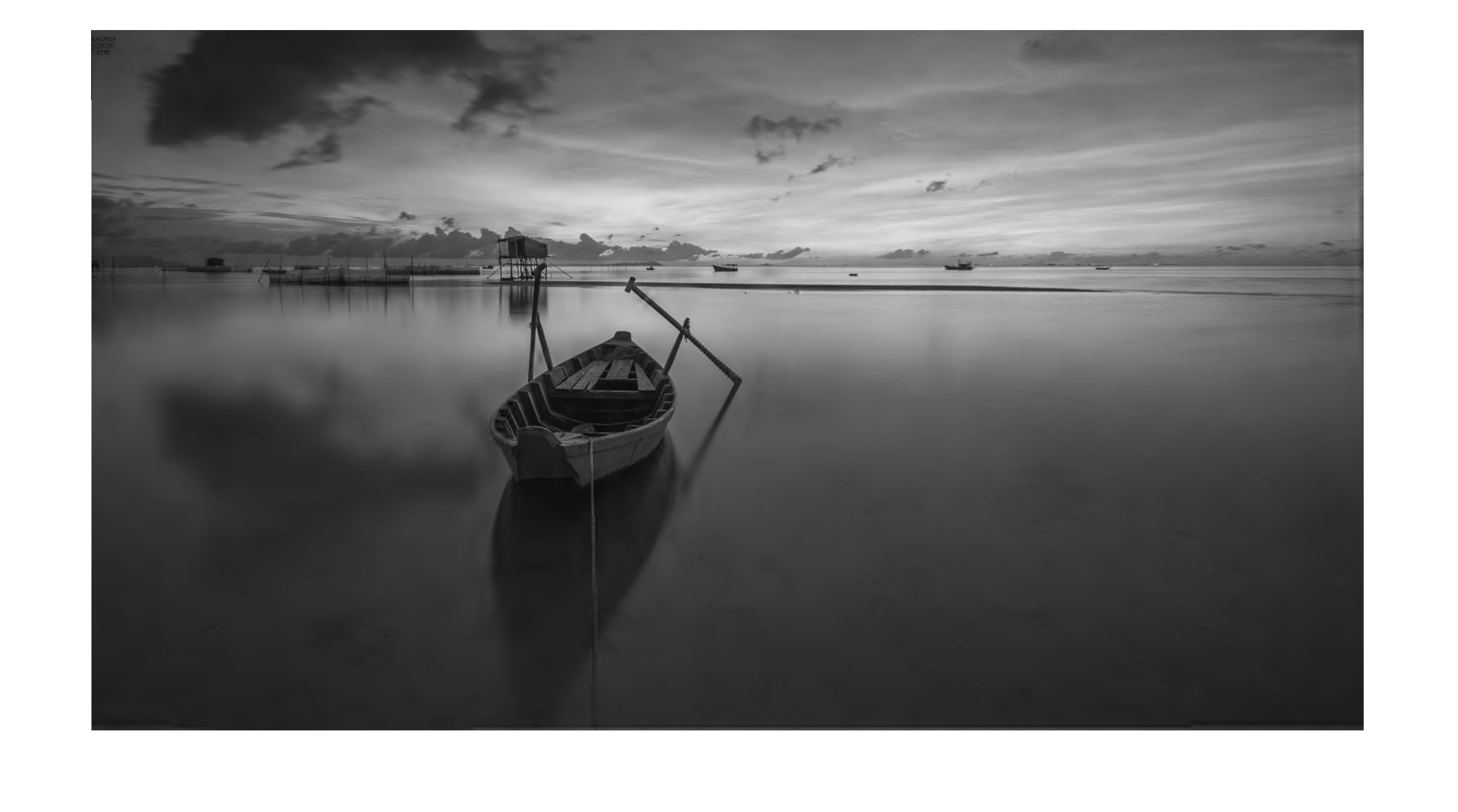}} \hfill
 \subfigure[Rate= 0.125 bpp]{\includegraphics[width=0.32\columnwidth, height=2.5cm]{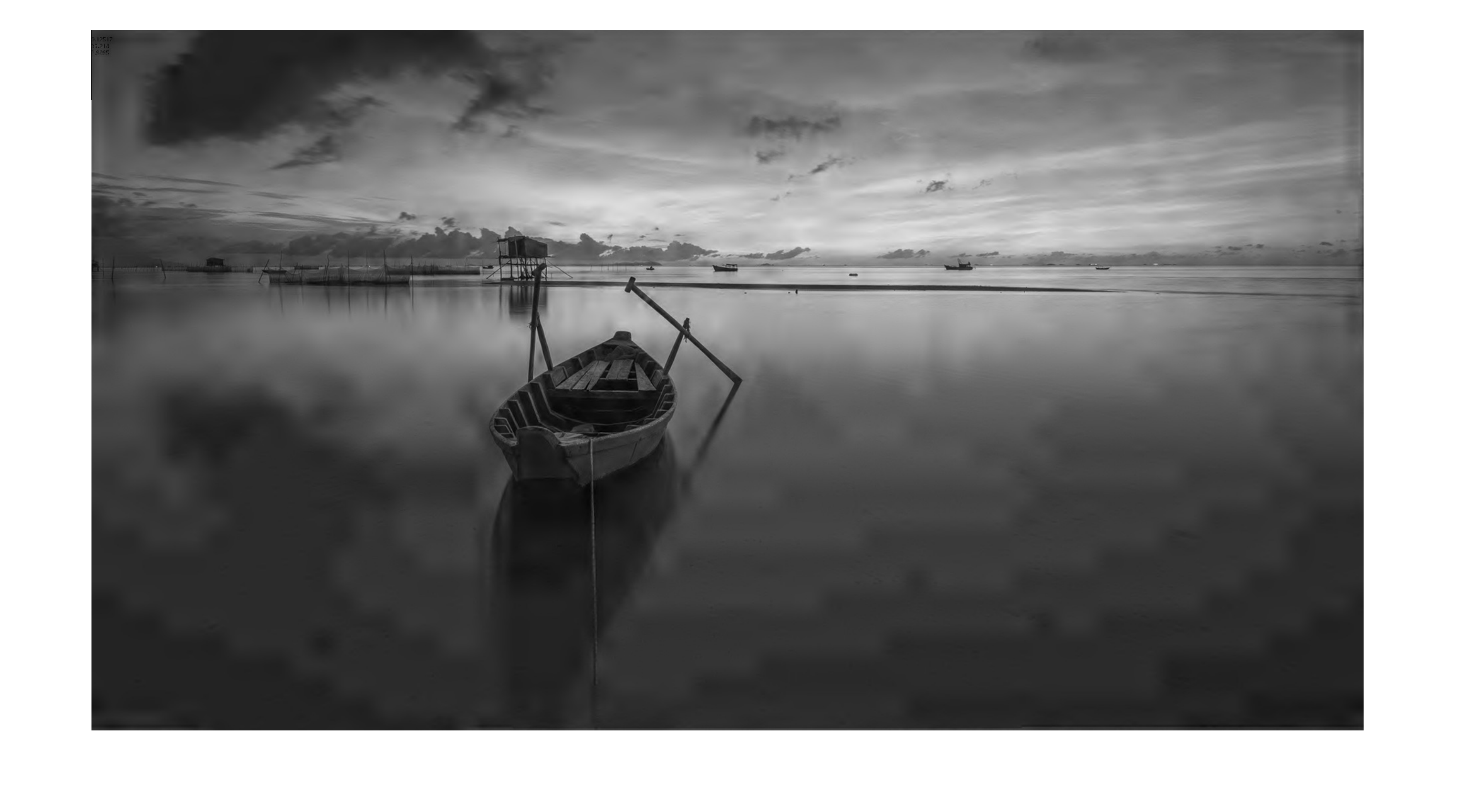}} \hfill
 \subfigure[Rate= 0.25 bpp]{\includegraphics[width=0.32\columnwidth, height=2.5cm]{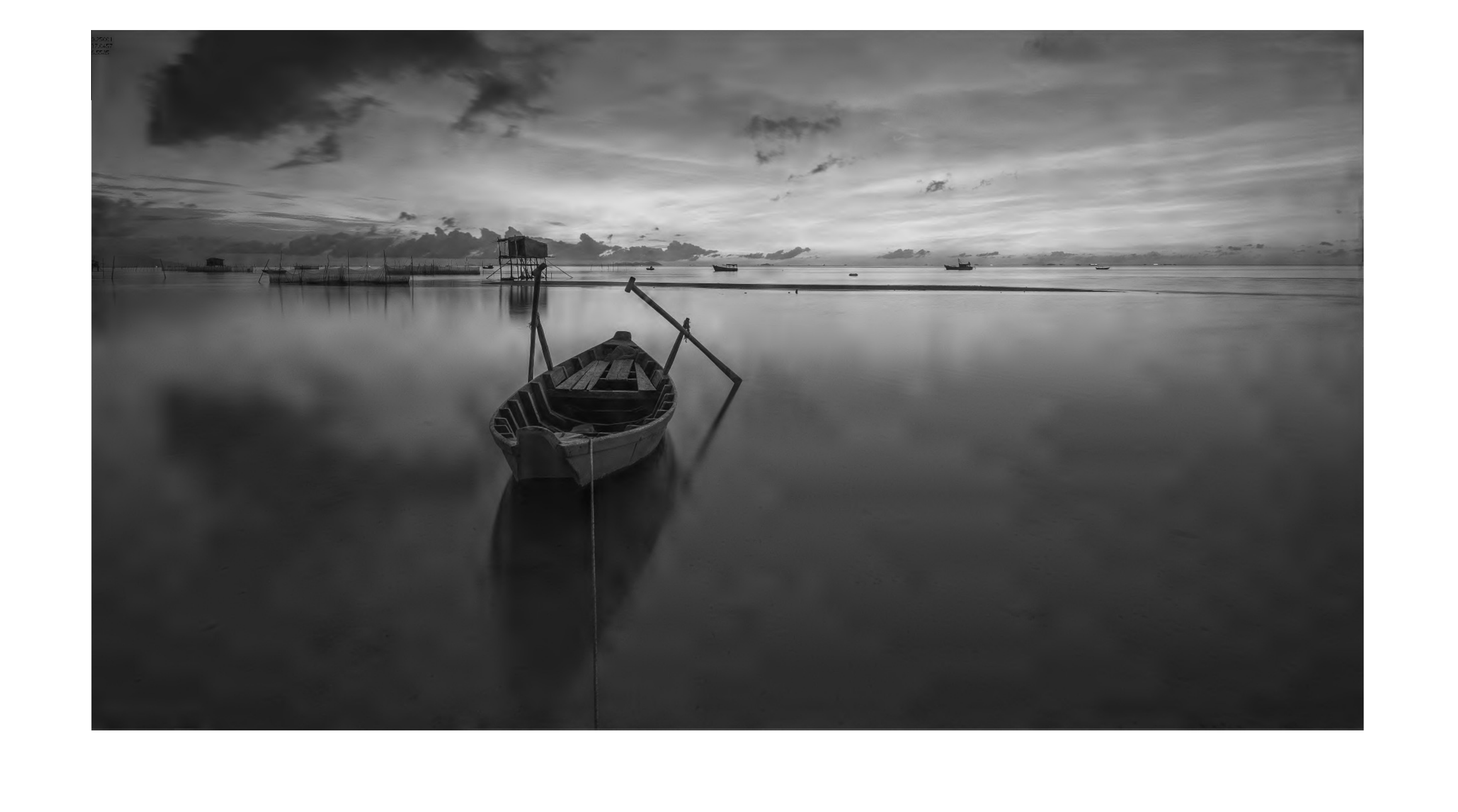}} \hfill
 \subfigure[Rate= 0.5 bpp]{\includegraphics[width=0.32\columnwidth, height=2.5cm]{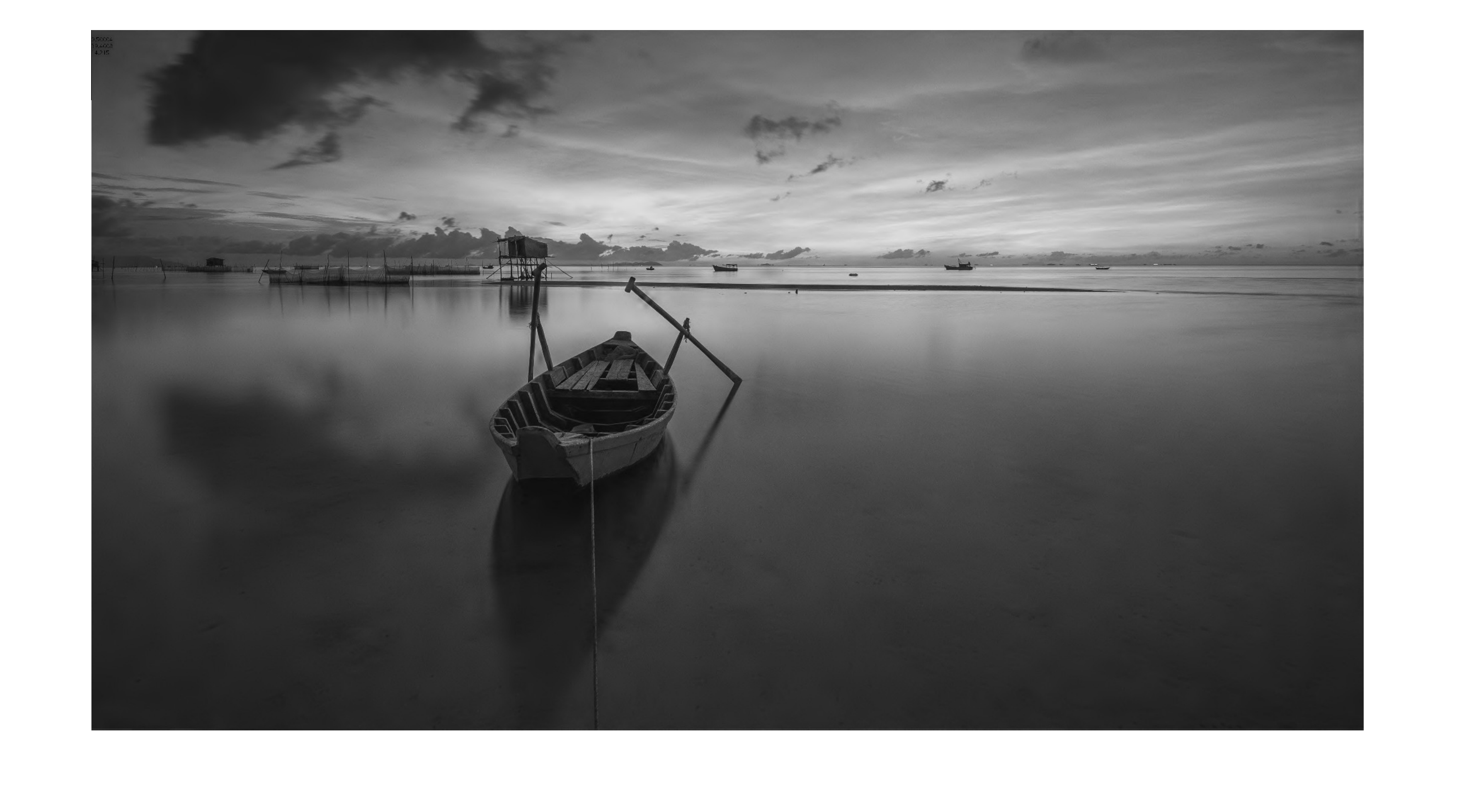}} \hfill
 \subfigure[Rate= 1.0 bpp]{\includegraphics[width=0.32\columnwidth, height=2.5cm]{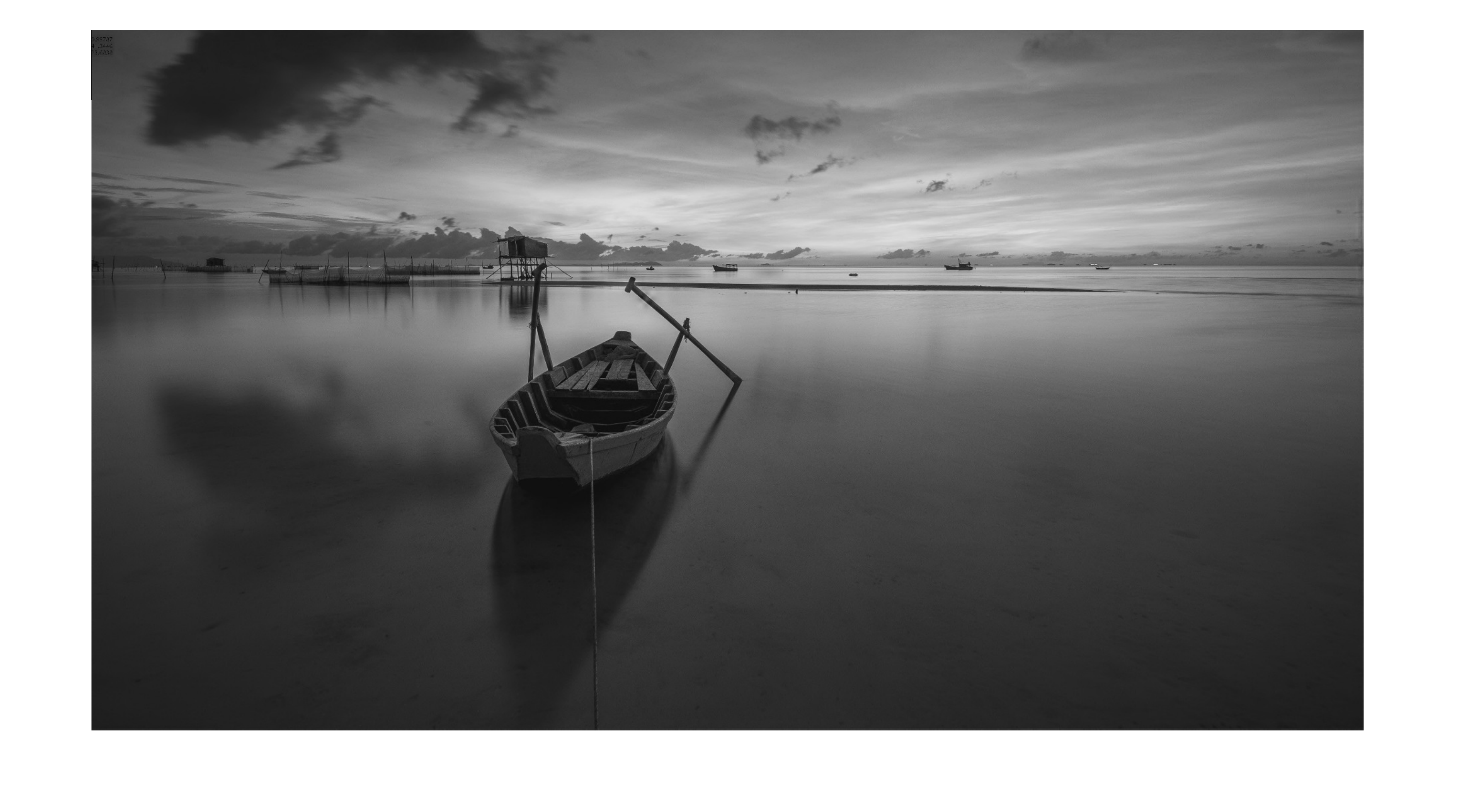}} \hfill
 \subfigure[Rate= 1.38 bpp]{\includegraphics[width=0.32\columnwidth, height=2.5cm]{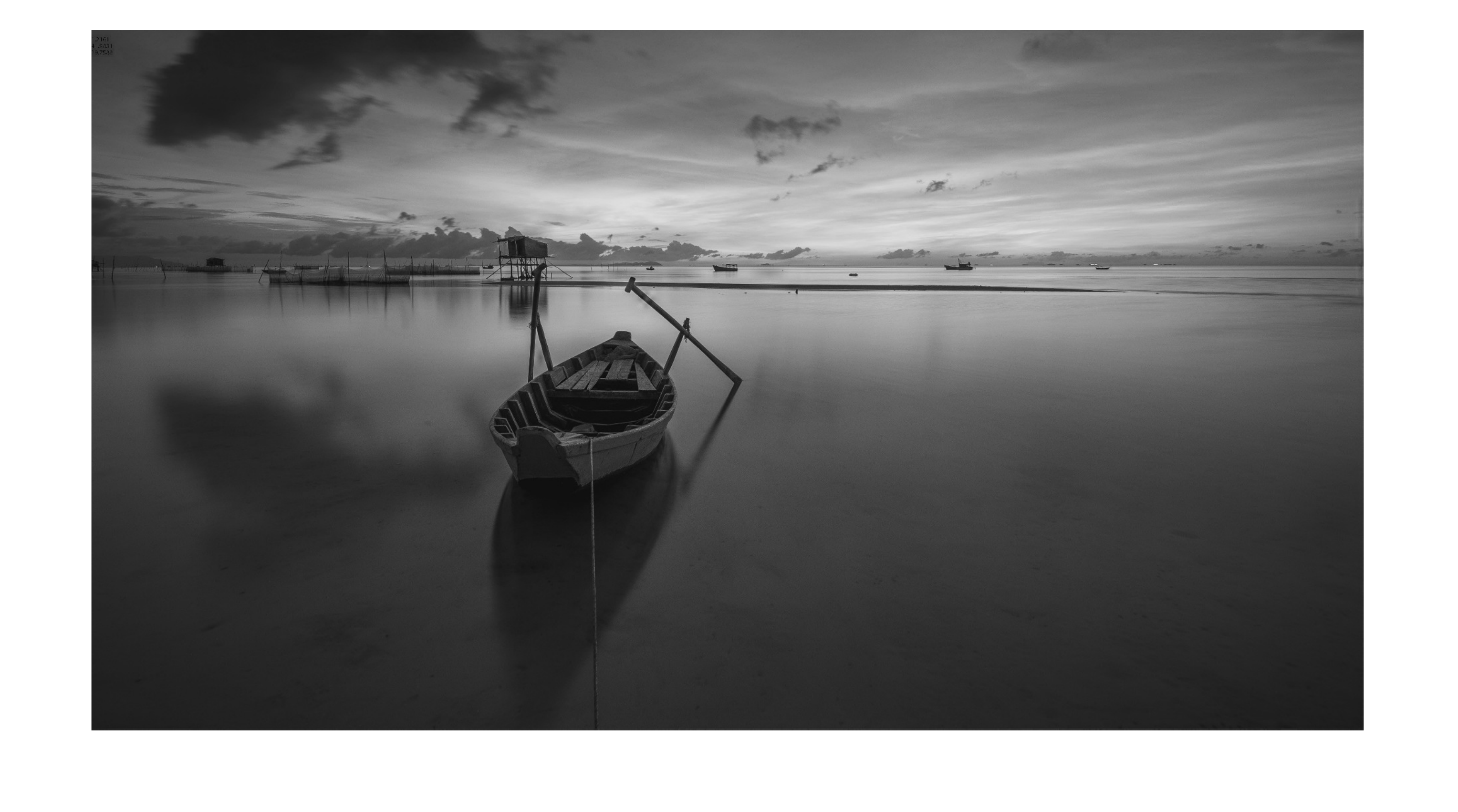}} \hfill
 \caption{Boat image reconstructed at standard rates: (a) Original Image, (b)-(g) Reconstructed at specified rates using the proposed BlinQS algorithm}
 \label{fig:boat_rates}
\end{figure*}

\subsection{Tabular and graphical results}
To evaluate the performance of the BlinQS under same platform, results have been compared against JPEG-2000 at standard rates (bpp) with and without quality layer in Table-\ref{table:compstdrates}. BlinQS has a clear domination over JPEG-2000 without quality layers and a near optimal value with quality layers. This shows that the estimation of BlinQS in optimizing the quality is very good. In JPEG-2000, the variation in PSNR is around 10dB for layered and non-layered string \cite{auli2008jpeg2000}, but using BlinQS that has been reduced by a large extent and satisfactory results in terms of visual quality and PSNR are obtained.

To further investigate the proposed method, test images with various resolutions and textures have been selected for comparison and some of the sample test images are presented in Table- \ref{table:sampledb}. The detailed PSNR values of around 100 images are presented in Appendix \ref{A1}. PSNR and Structural Similarity (SSIM) index values obtained for these images, using BlinQS and JPEG-2000 are presented in Tables-\ref{table:comptabledb_psnr} and \ref{table:comptabledb_ssim} respectively. From the PSNR values presented in Table- \ref{table:comptabledb_psnr}, it can be observed that even at lower rates BlinQS is giving more quality than JPEG-2000 without quality layer (J2K$^\#$) and PSNR is $\approx$30dB (but $<$J2K value) which clearly tells the visual quality of the image is flawless. From the SSIM values presented in Table- \ref{table:comptabledb_ssim}, it can be clearly seen that BlinQS is giving almost same results as that of J2K and giving very goos results when compared to J2K$^\#$. he comparison graph between ``PSNR (dB) and Rate'' of Lena (512$\times$512) and Barbara (512$\times$512) is depicted in Figure \ref{fig:psnrvsrate}. From this graph, it can be clearly seen that BlinQS is almost following JPEG-2000 in-terms of PSNR value. It is even closer at lower rates like 0.0625 bpp and 0.125 bpp when compared to other rates.

\begin{sidewaystable}[htbp]
  \centering
  \caption{PSNR (dB) comparison of BlinQS and JPEG-2000 for sample database}
  \resizebox{\columnwidth}{!}{
    \begin{tabular}{crrrrrrrrrrrrrrr}
    \toprule\noalign{\smallskip}
    \textbf{Rate} & \multicolumn{3}{c}{\textbf{0.0625 bpp}} & \multicolumn{3}{c}{\textbf{0.125 bpp}} & \multicolumn{3}{c}{\textbf{0.25 bpp}} & \multicolumn{3}{c}{\textbf{0.5 bpp}} & \multicolumn{3}{c}{\textbf{1 bpp}} \\\noalign{\smallskip}
    \midrule\noalign{\smallskip}
          \textbf{S. No} & \multicolumn{1}{c}{\textbf{J2K}} & \multicolumn{1}{l}{\textbf{J2K$^\#$}} & \multicolumn{1}{c}{\textbf{BlinQS}} & \multicolumn{1}{c}{\textbf{J2K}} & \multicolumn{1}{l}{\textbf{J2K$^\#$}} & \multicolumn{1}{c}{\textbf{BlinQS}} & \multicolumn{1}{c}{\textbf{J2K}} & \multicolumn{1}{l}{\textbf{J2K$^\#$}} & \multicolumn{1}{c}{\textbf{BlinQS}} & \multicolumn{1}{c}{\textbf{J2K}} & \multicolumn{1}{l}{\textbf{J2K$^\#$}} & \multicolumn{1}{c}{\textbf{BlinQS}} & \multicolumn{1}{c}{\textbf{J2K}} & \multicolumn{1}{l}{\textbf{J2K$^\#$}} & \multicolumn{1}{c}{\textbf{BlinQS}} \\\noalign{\smallskip}
    \midrule\noalign{\smallskip}
    1     & 20.69 & 19.17 & 20.38 & 21.69 & 19.84 & 20.94 & 23.15 & 19.99 & 21.16 & 25.48 & 20.90 & 22.87 & 28.97 & 21.79 & 26.11 \\\noalign{\smallskip}
    2     & 26.17 & 20.48 & 24.84 & 29.31 & 22.62 & 26.41 & 32.55 & 24.35 & 28.43 & 36.58 & 26.75 & 33.40 & 41.25 & 30.38 & 37.57 \\\noalign{\smallskip}
    3     & 27.69 & 21.49 & 25.39 & 30.83 & 24.35 & 27.45 & 33.44 & 24.96 & 29.80 & 35.73 & 28.70 & 32.76 & 38.20 & 31.66 & 36.23 \\\noalign{\smallskip}
    4     & 25.21 & 21.01 & 24.19 & 27.41 & 22.79 & 25.51 & 30.02 & 23.02 & 27.18 & 33.20 & 25.97 & 29.95 & 36.64 & 28.96 & 34.50 \\\noalign{\smallskip}
    5    & 39.64 & 34.34 & 36.54 & 40.34 & 34.97 & 35.29 & 41.13 & 36.55 & 37.08 & 42.53 & 37.36 & 39.77 & 44.88 & 39.66 & 41.05 \\\noalign{\smallskip}
    6    & 39.64 & 28.7  & 36.54 & 40.34 & 29.10 & 35.29 & 41.13 & 29.80 & 37.08 & 42.53 & 30.09 & 39.77 & 44.88 & 31.20 & 41.05 \\\noalign{\smallskip}
    \midrule\noalign{\smallskip}
    \textbf{Average} & 29.84 & 24.20 & 27.98 & 31.65 & 25.61 & 28.48 & 33.57 & 26.45 & 30.12 & 36.01 & 28.30 & 33.09 & 39.14 & 30.61 & 36.08 \\\noalign{\smallskip}
    \bottomrule
    \end{tabular}%
    }
  \label{table:comptabledb_psnr}%
  
\end{sidewaystable}%

% Table generated by Excel2LaTeX from sheet 'Sheet1 (2)'
\begin{sidewaystable}[htbp]
  \centering
  \caption{SSIM comparison of BlinQS and JPEG-2000 for sample database}
  \resizebox{\columnwidth}{!}{
    \begin{tabular}{crrrrrrrrrrrrrrr}
    \toprule\noalign{\smallskip}
    \textbf{Rate} & \multicolumn{3}{c}{\textbf{0.0625 bpp}} & \multicolumn{3}{c}{\textbf{0.125 bpp}} & \multicolumn{3}{c}{\textbf{0.25 bpp}} & \multicolumn{3}{c}{\textbf{0.5 bpp}} & \multicolumn{3}{c}{\textbf{1 bpp}} \\\noalign{\smallskip}
    \midrule\noalign{\smallskip}
        \textbf{S. No} & \multicolumn{1}{c}{\textbf{J2K}} & \multicolumn{1}{l}{\textbf{J2K$^\#$}} & \multicolumn{1}{c}{\textbf{BlinQS}} & \multicolumn{1}{c}{\textbf{J2K}} & \multicolumn{1}{l}{\textbf{J2K$^\#$}} & \multicolumn{1}{c}{\textbf{BlinQS}} & \multicolumn{1}{c}{\textbf{J2K}} & \multicolumn{1}{l}{\textbf{J2K$^\#$}} & \multicolumn{1}{c}{\textbf{BlinQS}} & \multicolumn{1}{c}{\textbf{J2K}} & \multicolumn{1}{l}{\textbf{J2K$^\#$}} & \multicolumn{1}{c}{\textbf{BlinQS}} & \multicolumn{1}{c}{\textbf{J2K}} & \multicolumn{1}{l}{\textbf{J2K$^\#$}} & \multicolumn{1}{c}{\textbf{BlinQS}} \\\noalign{\smallskip}
    \midrule\noalign{\smallskip}
    1     & 0.57  & 0.33  & 0.57  & 0.69  & 0.46  & 0.63  & 0.79  & 0.50  & 0.62  & 0.88  & 0.69  & 0.74  & 0.95  & 0.81  & 0.88 \\\noalign{\smallskip}
    2     & 0.84  & 0.62  & 0.80  & 0.91  & 0.75  & 0.85  & 0.95  & 0.85  & 0.88  & 0.97  & 0.93  & 0.95  & 0.99  & 0.98  & 0.99 \\\noalign{\smallskip}
    3     & 0.86  & 0.65  & 0.79  & 0.92  & 0.80  & 0.86  & 0.95  & 0.83  & 0.90  & 0.97  & 0.94  & 0.93  & 0.98  & 0.97  & 0.98 \\\noalign{\smallskip}
    4     & 0.76  & 0.52  & 0.72  & 0.84  & 0.65  & 0.77  & 0.92  & 0.67  & 0.82  & 0.96  & 0.87  & 0.90  & 0.98  & 0.94  & 0.97 \\\noalign{\smallskip}
    5    & 1.00  & 0.99  & 0.99  & 1.00  & 0.99  & 0.97  & 1.00  & 0.99  & 0.98  & 1.00  & 0.99  & 1.00  & 1.00  & 0.99  & 1.00 \\\noalign{\smallskip}
    6    & 1.00  & 0.96  & 0.99  & 1.00  & 0.99  & 0.97  & 1.00  & 0.99  & 0.98  & 1.00  & 0.99  & 1.00  & 1.00  & 0.99  & 1.00 \\\noalign{\smallskip}
    \midrule\noalign{\smallskip}
    \textbf{Average} & 0.84  & 0.68  & 0.81  & 0.89  & 0.77  & 0.84  & 0.93  & 0.81  & 0.86  & 0.96  & 0.90  & 0.92  & 0.98  & 0.95  & 0.97 \\\noalign{\smallskip}
    \bottomrule
    \end{tabular}%
    }
  \label{table:comptabledb_ssim}%
  
  \flushright{\footnotesize{*J2K: JPEG-2000, J2K$^\#$: JPEG-2000 without quality layers (\cite{kakadu2000})}} \hspace{2cm}
\end{sidewaystable}%

\subsection{Visual quality representation}
For qualitative analysis, reconstructed images of BlinQS at various standard rates have been presented in Figures-\ref{fig:lena_rates}, \ref{fig:barbara_rates} and \ref{fig:boat_rates}. Where, sub-figure (a), represents the original image used for encoding and (b) to (g) represents the images reconstructed at standard rates as mentioned in the figure captions. The maximum possible rate that can be obtained through the compressed string is represented in subfigure (g). The visual quality is flawless when observed at rates $R_{std}>$0.5, and quite good even for lower rates. PSNR values for the respective rates are given in Table-\ref{table:compstdrates} and also mentioned in the figure along with the obtained rate. Hence, it can be clearly seen that in both qualitative and quantitative analysis BlinQS has provided nearly same results as that of JPEG-2000.

\subsection{Computational complexity and trade-off}
The basic idea of BlinQS is to reduce the computational load by removing the iterative R-D optimization algorithm and achieve blind scalbility. The computational complexity of R-D optimization algorithm is given by $\mathcal{O}(N_{crv} \times N_{pt})$, where $N_{crv}$ represents the number of code-blocks and $N_{pt}$ represents number of average points in each code-block \cite{aminlou2006non}. Hence, BlinQS has reduced the computational load by the order of $\mathcal{O}(N_{crv} \times N_{pt})$ at a sacrifice of $\approx7\%$ of PSNR compared to JPEG-2000.

To achieve optimal quality, non-iterative and computationally less complex algorithm using gaussian normal distribution has been added to the decoder for standard rates. For non-standard rates, iterations are used to achieve optimal quality. It adds computational complexity of $\mathcal{O}(N_{depth})$, where $N_{depth}$ represents precision of the target rate that depends on the threshold and target rate selected by the user. This has also added a \emph{new degree of freedom} for choosing any required rate at the decoder rather than limiting to the rates calculated at the encoder. Hence, loss of $\approx7\%$ of PSNR has resulted in the decoder independency for optimal reconstruction at target rate and reduced computational load.

\section{Conclusion} \label{sec:conclusion}
This paper addresses the necessity of blind quality scalability for image compression and its implementation. The main concerns of quality scalability, iterative coding and lack of scalability for single layered string, are taken into consideration for developing BlinQS (Blind Quality Scalable) image compression. Normal distribution of percentage lengths has been used for getting the inclusion map for the target rate and this map is used for generating the truncation points $(n_i)$ for the respective blocks. Algorithm to obtain the inclusion map for achieving optimum reconstruction quality without iterative process at decoder has been introduced. This has reduced the computational complexity by a factor of $\mathcal{O}(N_{crv} \times N_{pt})$. The PSNR values obtained by the proposed algorithm have been presented in the the comparison table, which shows that BlinQS has obtained nearly same results using single string without using quality layers. Results shown for standard images clearly show the visual quality of the reconstructed image is flawless at higher rates and quite good even at lower rates. On an average, PSNR values obtained for BlinQS are 7\% less than that of JPEG-2000 by reducing the computational load on encoder and making the single string scalable at any desired target rate.

% \ifCLASSOPTIONcaptionsoff
%   \newpage
% \fi
% \section*{Funding}
% No funding have been received from external sources. Research is purely done as a part of academic work.
% 
% \section*{Conflicts of interest/ Competing interests}
% No conflicts on interst.
% 
% \section*{Availability of data and material}
% Datasets would be made available on the request.
% 
% \section*{Code Availability}
% Code is not avaialble on any public domain.

\addcontentsline{toc}{chapter}{REFERENCES}
\bibliographystyle{ieeetr}
\bibliography{referenceblinqs}

\section*{Appendix} \label{A1}
This appendix presents the comparison results of BlinQS algorithm against JPEG-2000 with and without quality layers in Table \ref{tab:blinqsresults}. Images presented here comprise of standard images and other images downloaded from standard datasets like SCIEN test images and videos \cite{scien}, Laboratory for Image \& Video Engineering, Texas \cite{sheikh2005live}, \cite{wang2004image}, \cite{sheikh2006statistical}, Robert Chung colour management database and Roger K. Fawcett Distinguished Professor \cite{RobertChung}. In Table \ref{tab:blinqsresults},  J2K$^\#$ stands for JPEG-2000 without quality layers and J2K stands for JPEG-2000 with quality layers.
% \newpage
%\vfill
%\vspace{-20pt}
\begin{table}[H]
\caption{PSNR (dB)comparison of BlinQS against JPEG-2000 with and without quality layers at different rates (bpp)}
\label{tab:blinqsresults}
\centering
\resizebox{\columnwidth}{!}{
\begin{tabular}{p{0.1\textwidth}llrrrrr}
    %\centering
  
%   \endfirsthead
%   \endhead
   \hline
    \textbf{Image} & \multicolumn{1}{l}{\textbf{Method}} & \multicolumn{5}{c}{\textbf{PSNR for different rates}} \\
    \textbf{Name} &       & \textbf{0.0625} & \textbf{0.125} & \textbf{0.25} & \textbf{0.5} & \textbf{1} \\
    \midrule
    \multirow{3}{*}{1.pgm} & J2K   & 23.37 & 25.42 & 28.30 & 32.08 & 37.08 \\
          & J2K$^\#$ & 20.08 & 22.14 & 22.89 & 23.78 & 24.92 \\
          & BlinQS & 22.64 & 23.63 & 25.17 & 28.54 & 34.36 \\
    \midrule
    \multirow{3}{*}{109.pgm}  & J2K   & 20.69 & 21.68 & 23.12 & 25.47 & 28.96 \\
          & J2K$^\#$ & 19.17 & 19.84 & 19.99 & 20.90 & 21.79 \\
          & BlinQS & 20.37 & 20.94 & 21.16 & 22.72 & 26.06 \\
    \midrule
    \multirow{3}{*}{111.pgm} & J2K   & 26.11 & 29.29 & 32.53 & 36.54 & 41.26 \\
          & J2K$^\#$ & 20.48 & 22.62 & 24.35 & 26.75 & 30.38 \\
          & BlinQS & 24.82 & 26.40 & 28.42 & 33.01 & 37.74 \\
    \midrule 
    \multirow{3}{*}{113.pgm} & J2K   & 27.69 & 30.81 & 33.43 & 35.72 & 38.19 \\
          & J2K$^\#$ & 21.49 & 24.35 & 24.96 & 28.70 & 31.66 \\
          & BlinQS & 25.39 & 27.43 & 29.78 & 31.96 & 36.12 \\
    \midrule
    \multirow{3}{*}{198.pgm}  & J2K   & 25.21 & 27.41 & 30.02 & 33.20 & 36.64 \\
          & J2K$^\#$ & 21.01 & 22.79 & 23.02 & 25.97 & 28.96 \\
          & BlinQS & 24.19 & 25.51 & 27.18 & 29.51 & 34.50 \\
    \midrule
    \multirow{3}{*}{N1A.pgm}   & J2K   & 23.48 & 25.02 & 27.27 & 30.66 & 35.72 \\
          & J2K$^\#$ & 21.47 & 21.78 & 22.10 & 22.51 & 24.33 \\
          & BlinQS & 21.97 & 22.58 & 23.12 & 25.31 & 30.37 \\
    \midrule
    \multirow{3}{*}{N5A.pgm}   & J2K   & 21.41 & 23.71 & 26.78 & 30.64 & 35.51 \\
          & J2K$^\#$ & 16.69 & 17.70 & 18.53 & 20.07 & 22.47 \\
          & BlinQS & 18.66 & 19.23 & 20.30 & 22.25 & 28.15 \\
    \midrule
    \multirow{3}{*}{building2.bmp} &   J2K   & 17.79 & 19.14 & 20.92 & 23.42 & 27.30 \\
          & J2K$^\#$ & 15.46 & 16.06 & 16.78 & 18.45 & 20.14 \\
          & BlinQS & 15.26 & 16.74 & 16.95 & 18.27 & 21.37 \\
    \midrule
    \multirow{3}{*}{buildings.bmp} &   J2K   & 19.44 & 21.09 & 23.43 & 26.75 & 31.49 \\
          & J2K$^\#$ & 15.99 & 17.42 & 18.20 & 19.74 & 21.71 \\
          & BlinQS & 18.67 & 19.25 & 20.71 & 22.31 & 27.14 \\
    \midrule
    \multirow{3}{*}{coins.bmp} &   J2K   & 23.17 & 24.93 & 27.25 & 29.98 & 34.18 \\
          & J2K$^\#$ & 19.42 & 21.35 & 21.83 & 24.45 & 26.29 \\
          & BlinQS & 22.28 & 21.09 & 22.24 & 23.89 & 29.23 \\
    \midrule
    \multirow{3}{*}{elaine.pgm}  & J2K   & 29.34 & 31.14 & 32.33 & 33.52 & 36.07 \\
          & J2K$^\#$ & 23.15 & 27.28 & 28.62 & 30.88 & 32.66 \\
          & BlinQS & 27.06 & 27.85 & 28.70 & 30.69 & 34.38 \\
    \midrule
    \multirow{3}{*}{fhd2.pgm}   & J2K   & 23.63 & 25.20 & 27.19 & 30.05 & 34.51 \\
          & J2K$^\#$ & 20.83 & 22.04 & 22.84 & 24.36 & 25.75 \\
          & BlinQS & 22.90 & 22.39 & 22.72 & 24.71 & 30.19 \\
    \midrule
    \multirow{3}{*}{fhd3.pgm}   & J2K   & 43.26 & 47.37 & 50.52 & 52.65 & 54.37 \\
          & J2K$^\#$ & 32.67 & 37.62 & 42.27 & 48.19 & 54.40 \\
          & BlinQS & 35.82 & 38.61 & 42.28 & 46.63 & 46.63 \\
    \midrule   
    \multirow{3}{*}{fhd4.pgm}   & J2K   & 28.73 & 31.45 & 34.53 & 37.94 & 42.21 \\
          & J2K$^\#$ & 22.94 & 25.88 & 28.19 & 31.72 & 35.54 \\
          & BlinQS & 24.39 & 23.16 & 25.62 & 29.16 & 38.78 \\
    \midrule
    \multirow{3}{*}{flowers.bmp} &   J2K   & 18.39 & 19.65 & 21.47 & 24.05 & 28.50 \\
          & J2K$^\#$ & 16.18 & 16.98 & 17.50 & 18.88 & 20.27 \\
          & BlinQS & 15.55 & 16.41 & 17.82 & 19.80 & 24.29 \\
    \midrule
     \multirow{3}{*}{img2.pgm}  & J2K   & 34.19 & 36.93 & 39.14 & 41.08 & 43.49 \\
          & J2K$^\#$ & 25.42 & 28.15 & 29.62 & 32.25 & 37.43 \\
          & BlinQS & 30.82 & 33.49 & 36.25 & 38.92 & 40.01 \\
    \midrule
    \multirow{3}{*}{k01.bmp}   & J2K   & 29.40 & 32.06 & 35.40 & 38.88 & 42.37 \\
          & J2K$^\#$ & 23.08 & 25.42 & 26.82 & 30.87 & 35.65 \\
          & BlinQS & 25.91 & 28.02 & 30.99 & 34.26 & 35.26 \\
    \bottomrule 
\end{tabular}%
}
\end{table}

% \begin{table}
% \caption{PSNR (dB)comparison of BlinQS against JPEG-2000 with and without quality layers at different rates (bpp)-2}
% \centernig
\resizebox{\columnwidth}{!}{
\begin{tabular}{p{0.1\textwidth}llrrrrr}
    %\centering
  
%   \endfirsthead
%   \endhead
   \hline
    \textbf{Image} & \multicolumn{1}{l}{\textbf{Method}} & \multicolumn{5}{c}{\textbf{PSNR for different rates}} \\
    \textbf{Name} &       & \textbf{0.0625} & \textbf{0.125} & \textbf{0.25} & \textbf{0.5} & \textbf{1} \\
    
    \toprule
    \multirow{3}{*}{k06.bmp}   & J2K   & 32.18 & 35.28 & 38.73 & 41.86 & 45.04 \\
          & J2K$^\#$ & 25.47 & 27.85 & 29.65 & 33.55 & 39.66 \\
          & BlinQS & 29.14 & 32.29 & 35.67 & 40.55 & 40.77 \\
    \midrule
    \multirow{3}{*}{k08.bmp}   & J2K   & 27.45 & 29.86 & 32.22 & 34.27 & 36.90 \\
          & J2K$^\#$ & 21.23 & 23.60 & 25.66 & 28.30 & 31.52 \\
          & BlinQS & 25.23 & 26.54 & 29.15 & 31.81 & 35.52 \\
    \midrule
    \multirow{3}{*}{k09.bmp}  & J2K   & 34.78 & 35.40 & 36.01 & 37.11 & 38.96 \\
          & J2K$^\#$ & 29.61 & 31.45 & 33.14 & 34.08 & 36.68 \\
          & BlinQS & 33.58 & 34.65 & 35.19 & 36.15 & 36.73 \\
    \midrule
    \multirow{3}{*}{k10.bmp}  & J2K   & 35.04 & 35.74 & 36.43 & 37.58 & 39.59 \\
          & J2K$^\#$ & 29.83 & 32.13 & 33.65 & 35.60 & 37.19 \\
          & BlinQS & 33.88 & 34.96 & 35.59 & 36.71 & 37.18 \\
    \midrule
    \multirow{3}{*}{k12.bmp}   & J2K   & 39.51 & 41.53 & 43.11 & 44.53 & 46.74 \\
          & J2K$^\#$ & 31.52 & 34.15 & 37.69 & 39.35 & 42.51 \\
          & BlinQS & 36.22 & 38.16 & 40.18 & 42.29 & 42.29 \\
    \midrule
    \multirow{3}{*}{k13.bmp}   & J2K   & 26.14 & 28.34 & 31.17 & 34.65 & 38.91 \\
          & J2K$^\#$ & 21.96 & 23.72 & 24.90 & 28.45 & 30.20 \\
          & BlinQS & 23.86 & 24.50 & 26.71 & 31.30 & 37.15 \\
    \midrule
    \multirow{3}{*}{k15.bmp}   & J2K   & 37.01 & 39.54 & 42.05 & 44.52 & 47.39 \\
          & J2K$^\#$ & 29.84 & 32.32 & 35.36 & 37.95 & 41.83 \\
          & BlinQS & 33.43 & 35.64 & 38.37 & 42.02 & 42.02 \\
    \midrule
    \multirow{3}{*}{k16.bmp}   & J2K   & 34.02 & 36.71 & 39.45 & 42.07 & 44.61 \\
          & J2K$^\#$ & 27.23 & 29.51 & 30.72 & 33.83 & 39.44 \\
          & BlinQS & 31.93 & 34.72 & 37.33 & 40.61 & 40.61 \\
    \midrule
    \multirow{3}{*}{k19.bmp}  & J2K   & 33.63 & 36.06 & 38.44 & 40.62 & 43.69 \\
          & J2K$^\#$ & 26.61 & 28.69 & 31.05 & 34.18 & 39.98 \\
          & BlinQS & 30.95 & 33.15 & 36.10 & 39.41 & 39.70 \\
    \midrule
    \multirow{3}{*}{k20.bmp}   & J2K   & 36.20 & 38.41 & 40.16 & 41.85 & 44.29 \\
          & J2K$^\#$ & 29.19 & 31.42 & 34.46 & 36.59 & 41.06 \\
          & BlinQS & 32.64 & 34.82 & 36.57 & 37.85 & 37.85 \\
    \midrule
    \multirow{3}{*}{k21.bmp}   & J2K   & 32.08 & 34.47 & 36.65 & 38.68 & 41.52 \\
          & J2K$^\#$ & 25.36 & 28.16 & 29.08 & 32.78 & 37.17 \\
          & BlinQS & 29.37 & 31.52 & 33.92 & 37.41 & 38.56 \\
    \midrule
    \multirow{3}{*}{k22.bmp}   & J2K   & 33.90 & 36.28 & 38.71 & 41.07 & 44.42 \\
          & J2K$^\#$ & 28.72 & 30.67 & 32.84 & 35.97 & 40.38 \\
          & BlinQS & 31.18 & 33.59 & 36.30 & 39.90 & 40.03 \\
    \midrule
    \multirow{3}{*}{k23.bmp}   & J2K   & 40.14 & 41.31 & 42.53 & 44.36 & 46.99 \\
          & J2K$^\#$ & 33.16 & 36.10 & 39.89 & 41.72 & 44.46 \\
          & BlinQS & 35.76 & 36.71 & 37.52 & 37.89 & 37.89 \\
    \midrule
     \multirow{3}{*}{lena512.pgm}  & J2K   & 28.15 & 31.01 & 33.89 & 37.08 & 40.27 \\
          & J2K$^\#$ & 22.31 & 25.29 & 27.30 & 29.45 & 33.13 \\
          & BlinQS & 22.26 & 25.01 & 28.91 & 33.97 & 38.34 \\
	\midrule   
    \multirow{3}{*}{lighthouse.bmp} &   J2K   & 24.72 & 26.38 & 28.59 & 31.98 & 37.12 \\
          & J2K$^\#$ & 20.96 & 22.40 & 23.06 & 24.48 & 26.60 \\
          & BlinQS & 23.94 & 24.38 & 25.35 & 27.97 & 33.55 \\
          \midrule
    \multirow{3}{*}{plane.bmp} &   J2K   & 28.41 & 30.70 & 33.33 & 37.12 & 43.08 \\
          & J2K$^\#$ & 23.07 & 24.54 & 26.69 & 28.09 & 31.36 \\
          & BlinQS & 25.86 & 26.99 & 28.17 & 31.60 & 35.43 \\
    \midrule
    \multirow{3}{*}{sailing1.bmp} &   J2K   & 24.41 & 25.68 & 27.80 & 30.88 & 35.51 \\
          & J2K$^\#$ & 21.20 & 22.40 & 22.84 & 24.31 & 26.50 \\
          & BlinQS & 23.75 & 24.26 & 25.30 & 27.23 & 31.40 \\
    \bottomrule

%     \begin{minipage}{1\columnwidth}
%     \hfill \flushright{\footnotesize{J2K: JPEG-2000 with quality layers from \cite{kakadu2000}, \cite{openjpegmaster},\\  J2K$^\#$- JPEG-2000 without quality layers from \cite{kakadu2000}, \cite{openjpegmaster}}}
%     \end{minipage}

\end{tabular}%
}
% \end{table}

\centering
% \begin{minipage}{0.45\textwidth}
\resizebox{\columnwidth}{!}{
\begin{tabular}{p{0.1\textwidth}llrrrrr}
    %\centering
  
%   \endfirsthead
%   \endhead
   \hline
    \textbf{Image} & \multicolumn{1}{l}{\textbf{Method}} & \multicolumn{5}{c}{\textbf{PSNR for different rates}} \\
    \textbf{Name} &       & \textbf{0.0625} & \textbf{0.125} & \textbf{0.25} & \textbf{0.5} & \textbf{1} \\
    \toprule
    \multirow{3}{*}{uhd.pgm}   & J2K   & 30.49 & 32.58 & 35.37 & 39.46 & 45.30 \\
          & J2K$^\#$ & 26.31 & 27.87 & 29.38 & 30.94 & 34.25 \\
          & BlinQS & 25.59 & 26.24 & 28.74 & 33.97 & 41.53 \\
    \midrule
    \multirow{3}{*}{uhd1.pgm}   & J2K   & 43.85 & 45.64 & 47.79 & 50.30 & 53.29 \\
          & J2K$^\#$ & 38.49 & 41.12 & 43.88 & 46.03 & 50.98 \\
          & BlinQS & 35.50 & 38.62 & 42.81 & 45.65 & 45.87 \\
    \midrule
    \multirow{3}{*}{uhd10.pgm}   & J2K   & 30.56 & 33.44 & 37.29 & 41.88 & 46.84 \\
          & J2K$^\#$ & 24.50 & 25.75 & 26.75 & 28.65 & 34.74 \\
          & BlinQS & 25.11 & 27.08 & 29.72 & 35.52 & 42.14 \\
    \midrule
    \multirow{3}{*}{uhd7.pgm}   & J2K   & 28.70 & 32.07 & 36.18 & 41.30 & 47.01 \\
          & J2K$^\#$ & 22.46 & 24.31 & 26.24 & 28.84 & 35.27 \\
          & BlinQS & 22.39 & 24.75 & 28.31 & 34.41 & 42.47 \\
    \midrule
    \multirow{3}{*}{uhd8.pgm}   & J2K   & 39.61 & 40.31 & 41.11 & 42.50 & 44.85 \\
          & J2K$^\#$ & 34.34 & 34.97 & 36.55 & 37.36 & 39.66 \\
          & BlinQS & 36.50 & 35.22 & 37.05 & 39.46 & 41.34 \\
    \midrule
    \multirow{3}{*}{uhd9.pgm}   & J2K   & 30.42 & 31.16 & 32.57 & 34.78 & 38.93 \\
          & J2K$^\#$ & 28.73 & 29.10 & 29.82 & 30.09 & 31.20 \\
          & BlinQS & 29.32 & 25.25 & 25.20 & 30.19 & 35.54 \\
    \midrule
    \multirow{3}{*}{us092.pgm} & J2K   & 20.02 & 21.26 & 23.37 & 26.48 & 31.40 \\
          & J2K$^\#$ & 17.44 & 18.46 & 19.13 & 20.05 & 21.15 \\
          & BlinQS & 17.49 & 18.80 & 17.95 & 20.42 & 25.76 \\
    \bottomrule

    \begin{minipage}{1\columnwidth}
    \hfill \flushright{\footnotesize{J2K: JPEG-2000 with quality layers from \cite{kakadu2000}, \cite{openjpegmaster},\\  J2K$^\#$- JPEG-2000 without quality layers from \cite{kakadu2000}, \cite{openjpegmaster}}}
    \end{minipage}

\end{tabular}%
}

\end{document}